\documentclass[fleqn,usenatbib]{mnras}

\usepackage{newtxtext,newtxmath}
\usepackage[T1]{fontenc}
\DeclareRobustCommand{\VAN}[3]{#2}
\let\VANthebibliography\thebibliography
\def\thebibliography{\DeclareRobustCommand{\VAN}[3]{##3}\VANthebibliography}
\usepackage{amsmath}	
\usepackage{graphicx}
\usepackage[dvipsnames]{xcolor}
\usepackage[normalem]{ulem}
\usepackage{threeparttable}
\usepackage{stackengine}	
\usepackage{url}
\usepackage{verbatim}
\usepackage[normalem]{ulem}
\usepackage{rotating}
\usepackage{bm}
\makeatletter
\let\@makecaption=\SFB@makefigurecaption
\makeatother

\newcommand{\ergs}{\rm{ erg \,s}^{-1}}

\newcommand{\kpc}{\rm~kpc}
\newcommand{\pc}{{\rm~pc}}

\newcommand{\flux}{\rm {erg\,s^{-1}\,Hz^{-1}\,kpc^{-2}\,sr^{-1}}}

\title[Non-thermal emission in jets and winds]{
Non-thermal emission in jets and winds:\\
Expected emission and spectral index distributions}

\author[Meenakshi et al.]{
M. Meenakshi$^{1,2}$\thanks{E-mail: mmeenakshi@aip.de}, D. Mukherjee$^{2}$\thanks{E-mail: dipanjan@iucaa.in}, G. Bodo$^{3}$, P. Rossi$^{3}$, C. M. Harrison$^{4}$, L. K. Morabito$^{5,6}$, \newauthor 
  P. Kharb$^{7}$ and S. Silpa$^{8}$
\\
$^{1}$Leibniz Institute for Astrophysics, An der Sternwarte 16, D-14482 Potsdam, Germany\\
$^{2}$IUCAA, Post Bag 4 Ganeshkhind, Savitribai Phule Pune University Campus Pune 411007, India\\
$^{3}$INAF, Osservatorio Astrofisico di Torino, Strada Osservatorio 20, I-10025 Pino Torinese, Italy\\
$^{4}$School of Mathematics, Statistics and Physics, Newcastle University, Newcastle upon Tyne NE1 7RU, UK\\
$^5$Centre for Extragalactic Astronomy, Department of Physics, Durham University, Durham DH1 3LE, UK\\
$^6$Institute for Computational Cosmology, Department of Physics, Durham University, Durham DH1 3LE, UK\\
$^7$National Centre for Radio Astrophysics (NCRA) - Tata Institute of Fundamental Research (TIFR), S. P. Pune University Campus, Pune 411007, India\\
$^8$Departamento de Astronom\'{i}a, Universidad de Concepci\'{o}n, Casilla 160-C, Concepci\'{o}n, Chile
}

\date{Accepted XXX. Received YYY; in original form ZZZ}

\pubyear{\the\year}

\begin{document}
\label{firstpage}
\pagerange{\pageref{firstpage}--\pageref{lastpage}}
\maketitle

\begin{abstract}

The origin of synchrotron emission in compact radio sources associated with active galactic nuclei (AGN) remains poorly understood. In a series of papers, we have examined diagnostic tools to disentangle the dominant underlying processes. In this study, we investigate the in situ evolution of cosmic-ray electrons (CREs) in compact AGN jets and winds, and examine how their evolution shapes the resulting observable radio properties. In jets, CREs experience multiple shock interactions as they propagate along the spine toward the hotspot and flow into the cocoon via backflows. In winds, CREs are predominantly accelerated at the Mach disc, with occasional re-acceleration within turbulent cocoon backflows. The continuous mixing of different CRE populations within the cocoon produces observational signatures that cannot be inferred from instantaneous conditions alone. In all jet simulations, spectral indices are flattest near the hotspot and steepen progressively away from the hotspots. In winds, spectra steepen with increasing distance from the Mach disc, with this trend becoming more pronounced at high radio frequencies due to radiative losses. We find the Mach disc to be a significantly more efficient CRE acceleration site than the forward shock in winds, which weakens as the wind expands to large scales. Since morphology, especially at low resolution, can be ambiguous for compact sources, spatially resolved spectral indices, particularly when combined with emission and polarization signatures, can provide a powerful diagnostic.

\end{abstract}

\section{Introduction}

Synchrotron emission from AGN jets has been recognized for decades as an important source for radio emission in active galaxies \citep[see][for a review]{saikia_2022}. With recent advances, however, synchrotron emission from AGN-driven winds has also emerged as a potentially significant contributor to the radio emission \citep{morabito_2019,petley_2022,petley_2024,fischer_2023,fawcett_2020,fawcett_2023,fawcett_2025}. In compact AGN, disentangling the emission from jets and winds is often challenging \citep{jarvis_2021,njeri_2025}, with observations suggesting that both mechanisms can sometimes operate simultaneously \citep{mehdipour_2019,silpa_2022,ghosh_2025,ghosh_2026}.

To address these challenges, our previous studies on compact ($\sim4$~kpc) AGN jets and winds \citep[][hereafter Paper~I and Paper~II, respectively]{meenakshi_2023, meenakshi_2024} examined the general characteristics of synchrotron emission and polarization. Those analyses, performed in post-processing, relied on instantaneous fluid properties (e.g., pressure, magnetic field strength) and assumed a fixed power-law energy spectrum for the cosmic-ray electrons (CREs). While this approach captured the global emission and polarization trends, revealing distinct signatures between jets and winds (see Figs.~13-15 in Paper~II), the differences in emission became less pronounced at lower resolutions. Moreover, this approach neglected the dynamical and radiative evolution of the electrons, which can substantially alter local spectra as the particles are advected, accelerated, and cooled throughout the flow.

To capture these processes more realistically, several recent numerical studies of AGN jets have adopted hybrid frameworks that evolve the non-thermal electron population self-consistently with the fluid \citep{jones_1999,mimica_2009,fromm_2016,mukherjee_2020,mukherjee_2021a,upreti_2024}. These approaches solve the phase-space evolution of the non-thermal electrons, including radiative losses and microphysical acceleration mechanisms such as diffusive shock acceleration \citep{drury_1983}. An alternative, yet related, strategy is adopted by other studies, which track Lagrangian tracer particles advected with the flow to model energy losses by the CREs in post-processing \citep{yates-jones_2022, yates-jones_2023, turner_2023,dubey_2023,dubey_2024,jerrim_2025, jlassi_2026}.


In jets, recollimation shocks and terminal hotspots serve as efficient acceleration sites, producing bright, multi-frequency emission from radio to X-rays \citep{clautice_2016,perlman_2019,He_2023}. While these features are prominent in extended ($\gtrsim$100 kpc) jets \citep[e.g.][]{bridle_1994}, their counterparts in compact, kiloparsec-scale jets remain comparatively elusive \citep{zuther_2012, McCaffrey_2022}. In contrast, the evolution of AGN-driven winds and the associated CRE populations has received relatively limited attention. Being wide-angled and mildly or non-relativistic, these winds evolve differently with the ambient medium compared to collimated jets \citep{meenakshi_2024}, leading to distinct emission characteristics. Theoretical studies have suggested that the termination shocks in AGN winds can be an important site to accelerate particles to high energies \citep[e.g.][]{bustard_2017, peretti_2023}, and a few and semi-analytical works have explored their radio signatures \citep{nims_2015, yamada_2024}. Nonetheless, the non-thermal properties and spectral evolution of AGN winds remain largely unexplored.

In this study, we build upon our previous works (Papers~I and II) by employing a Lagrangian approach to trace the evolution of CREs self-consistently. We use the \textsc{Lagrangian Particle} (LP) module implemented in the RMHD code \textsc{pluto} \citep{mignone_2007}, which enables a realistic treatment of electron transport, cooling, and acceleration \citep{vaidya_2018}. This enables us to compute the resulting multi-frequency synchrotron emission and spectral evolution of CREs, providing a more physically grounded understanding of how compact AGN jets and winds leave distinct imprints in the radio regime.

The paper~is structured as follows. In Sec.~\ref {sec:simulation-setup}, we describe the simulation setup, along with the details of CREs injection and evolution. Sec.~\ref {sec:results} presents the main results of the study, where we examine the evolution of CREs in Sec.~\ref{sec:cre_evolve_jw} and the corresponding updates in their spectra in Sec.~\ref{sec:join_spec_jw}. We then analyze the resulting multi-frequency emission and radio spectra from the compact jets and winds in Sec.~\ref {sec:synch_jw}, followed by a discussion of the emission characteristics of the large-scale jet in Sec.~\ref {sec:synch_j45}. In Sec.~\ref {sec:discussion}, we discuss the main findings of the paper and their connections to other studies. We conclude the key findings of the Paper~in Sec.~\ref{sec:conclusions}.

\begin{table*}
	\centering
	\caption{Parameters for the jet and wind simulations performed in this study.}
	\label{tab:sim_table}
	
	\begin{tabular}{|c|c|c|c|c|c|c|c|} 
		\hline
Simulation & Domain (${\rm kpc^3}$) &   $\,\eta_{j/w}$ & $\beta_{j/w}$ & $D_{j/w}$ ($\mathrm{pc}$) & $\mathrm{P}_{j/w} [\mathrm{\ergs}]$ &  $B_0 [\mathrm{mG}]$ & $N_T~(\times 10^6)$ \\
		\hline
			$\mathrm{J43}$ & 4 $\times$ 4 $\times$ 4  & $4{\rm E}{-5}$ & 0.866 & 100 & $1.25 {\rm E}{43}$ & 0.09 & 32\\
	\hline
	$\mathrm{W43-light}$ & 8 $\times$ 8 $\times$ 4 & $4{\rm E}{-2}$ & 0.051 & 400 & $1.03 {\rm E}{43}$ & 0.069 & 19 \\
		\hline
	$\mathrm{W43-dense}$ & 8 $\times$ 8 $\times$ 4 &  $4{\rm }{}$ &  0.011 & 400 & $1.06 {\rm E}{43}$ & 0.148 & 35 \\
			\hline

   $\mathrm{J44}$ & 4 $\times$ 4 $\times$ 4  & $4{\rm E}{-4}$ & 0.866 & 100 & $1.03 {\rm E}{44}$ & 0.26 & 13 \\
			\hline
            
   $\mathrm{W44-light}$ & 8 $\times$ 8 $\times$ 4  & $4{\rm E}{-2}$ & 0.11 &  400 & $1.06 {\rm E}{44}$ & 0.15 & 24 \\
			\hline

   $\mathrm{J45}$ & 6$\times$ 6 $\times$ 20  & $1{\rm E}{-3}$ & 0.98 & 200 & $1.2 {\rm E}{45}$ & 0.42 & 7.1 \\
		\hline
	\end{tabular}
	\begin{tablenotes}

\item Domain: Dimensions ($X \times Y \times Z$ in kpc$^3$) for the simulations performed. The jets and winds are launched along the $Z-$direction in all the cases.\\
$\eta_{j/w}$: Ratio of jet/wind density to the ambient gas density at the injection zone. \\
$\beta_{j/w}$: Z-component of the velocity for jet/wind in units of the speed of light.\\
$D_{j/w}$: Diameter at injection of the jet/wind.\\
 $\mathrm{P}_{j/w}$: Mechanical power of the jet/wind.\\
 $B_0 $: Maximum strength of the toroidal magnetic field of the jet/wind at the injection zone. \\
$N_T$: Total number of CREs in the domain at the last snapshots analysed in this study.
	\end{tablenotes}
	
\end{table*}

\begin{figure*}
   \centering
   \includegraphics[scale=0.65]{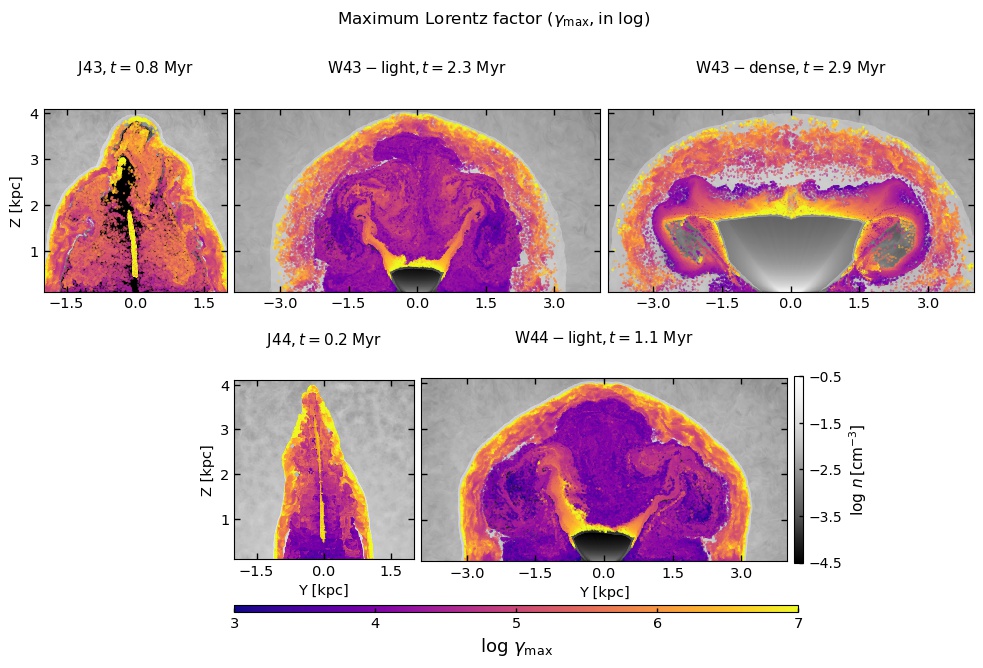}
      \caption{$Y-Z$ slices displaying Logarithmic maximum Lorentz factor for jets and winds of power $10^{43}~\ergs$ (top) and $10^{44}~\ergs$ (bottom) at different regions}. The times of the snapshot correspond to when the jets and winds have reached close to the upper $ Z$-boundary of the simulation box. CREs that have crossed at least one shock are shown. The background displays the logarithmic density, with darker areas indicating lower gas density. The time evolution for the first row can be seen at this \href{https://www.youtube.com/watch?v=s7mwNw_ce-g}{YouTube} link.
      
         \label{fig:lorentz_jw}
   \end{figure*}

\begin{figure*}
   \centering
   \includegraphics[scale=0.58]{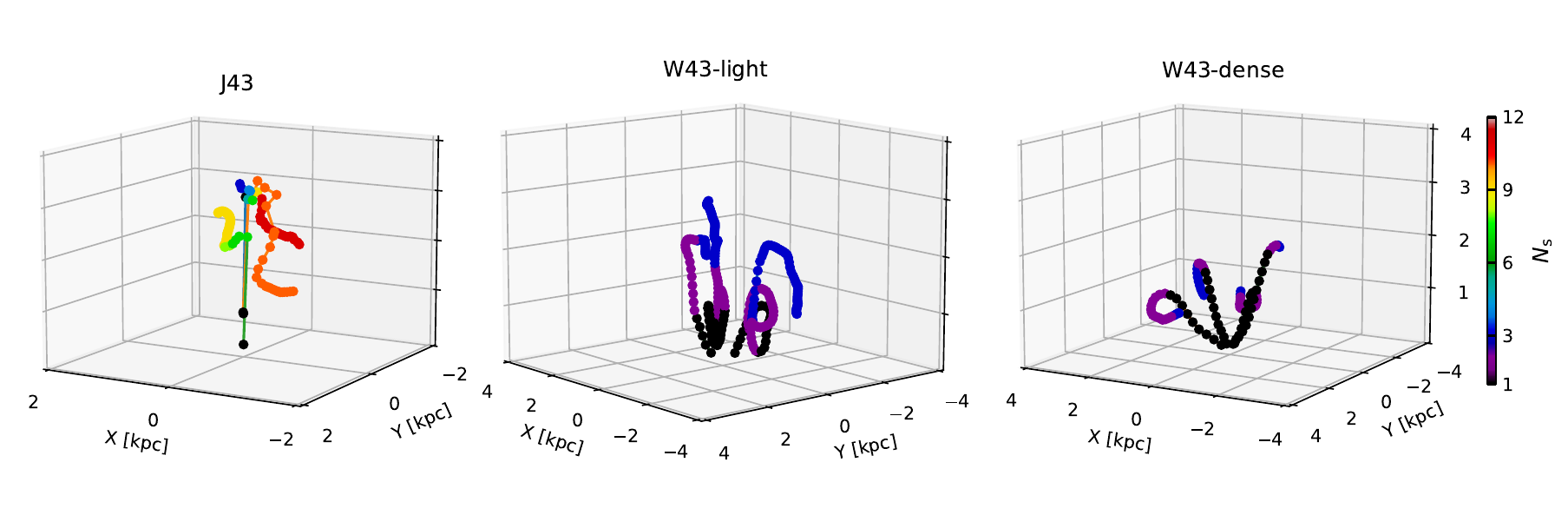}
      \caption{Sample trajectories of particles launched along the jet (J43) and winds (W43-light/dense). The color indicates the number of shocks ($N_{\rm s}$) crossed at a given point along the path.}
      
         \label{fig:shkn_jw}
   \end{figure*}

  \begin{figure}
   \centering
   \includegraphics[scale=0.6]{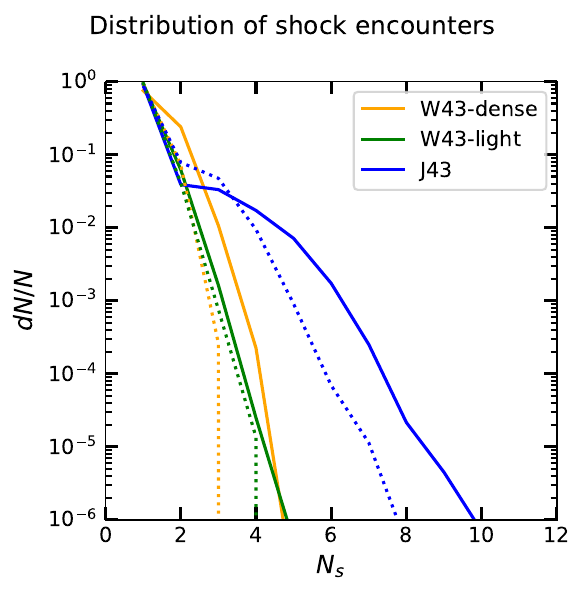}
      \caption{Distribution of shock crossings ($\geq 1$) experienced by CREs in jets and winds. Solid lines correspond to the times when the jets/winds reach the top of the simulation domain (see Fig.~\ref{fig:lorentz_jw}), while dotted lines represent half of those times. The distribution demonstrates that CREs in jets undergo a greater number of shock encounters compared to those in winds.}
         \label{fig:pdf_shkn}
   \end{figure}

\section{Simulation setup}
\label{sec:simulation-setup}
In this section, we briefly outline the simulation setup used for the jets and winds examined in this study. Further information on the overall setup and the (magneto)hydrodynamic framework can be found in Sec. 2 of Papers I and II. The ambient medium is set as a hot halo at a temperature of $10^7$~K, and its density and pressure values are estimated by assuming hydrostatic equilibrium in the external gravitational field ($\phi(r)$) as,
\begin{equation}
\frac{d p_a}{d r} = - \rho_a (r)\frac{d \phi (r)}{dr}; \qquad p_a(r) = \frac{\rho_a (r)}{\mu m_a} k_B T_h     
\end{equation}
from which we derive
\begin{equation}\label{eq:halo}
p_a(r) = (n_0 k_B T_h) \exp{\left[- \left( \frac{\mu m_a}{k_B T_h}\right) \phi (r) \right]}, T_h = 10^7\,{\rm K}   
\end{equation}
 where $p_a$ and $\rho_a = \mu m_a n_h$ are respectively the pressure and density of the halo gas at spherical radius $r$, $\mu=0.6$ is the mean molecular weight of the fully ionized gas, $m_a$ is the atomic weight, and $n_0$ is the number density at $r=0$. The Boltzmann constant is shown using $k_B$, and $T_h$ is the temperature of the halo. The total gravitational potential ($\phi(r)$) is the sum of the baryonic and dark-matter potential, as described in Paper~I (Sec. 2.1) in detail.

Throughout, we include a random (or turbulent) magnetic field distribution in the background (see Sec. 2.1.2 in Paper~I for numerical details), which is correlated over scales up to 1000~pc, as done in Paper~II for the winds. The boundary condition is set to outflow for all the boundaries of the domain, except the lower Z where it is reflective outside the injection region. Most of the runs are direct counterparts of our earlier work, with the jets and winds injected from ghost cells located at the lower $Z$-boundary of the computational domain. The simulations, with the injection parameters, are listed in Table~\ref{tab:sim_table}. We consider outflows with two mechanical powers, $10^{43}~\ergs$ and $10^{44}~\ergs$, representing low- to intermediate-power jets and winds. The injection parameters (density, pressure, velocities, and magnetic fields) are adjusted accordingly to achieve these energy inputs (see Sec. 2.2 in Paper I for more details). Similar to our previous works, the simulations are performed for one-sided jets/winds, which are launched with toroidal magnetic fields with a fixed magnetization\footnote{defined as the ratio of Poynting flux to the enthalpy flux} value of 0.1. The magnetic field follows a radial profile, which is tapered with a smoothing function outside the jet's radius. During injection, the $X$ and $Y$ components of the jet's magnetic field as a function of cylindrical radius ($r_c$) are estimated as:
\begin{equation}\label{eq:Bmag}
B_{(X,Y)} = \begin{cases}
\begin{matrix}
     B_{0} (r_c/R_j)  \quad \mathrm{for}\,\, r_c<=R_j\\
    B_{0} (r_c/R_j) \cosh{(r_c/R_j)}^6 \quad \mathrm{for}\,\, r_c>R_j
\end{matrix}
\end{cases}
\end{equation}
where $B_{0}$ are the peak strength values listed in Table~\ref{tab:sim_table} and $R_j$ is the jet's radius.

The compact jet runs, J43 and J44, correspond to powers of $10^{43}~\ergs$ and $10^{44}~\ergs$, respectively, each injected with a 100~pc diameter. These are analogous to the $\mathrm{J43\_L1000}$ and $\mathrm{J44\_L1000}$ runs from Paper~I. As shown in Table~\ref{tab:sim_table}, the higher-power jet has a greater density than the lower-power case, while maintaining a similar velocity. This implies that it injects roughly an order of magnitude more momentum flux into the system. The wind simulations: W43-light, W43-dense, and W44-light, represent light (low-density) and dense (high-density) winds with powers of $10^{43}~\ergs$ and $10^{44}~\ergs$, injected with a 400~pc diameter. These correspond to $\mathrm{W43\_L1000}$, $\mathrm{W43\_L1000HD}$, and $\mathrm{W44\_L1000}$ from Paper~II. The W43-dense wind has a density higher by roughly two orders of magnitude but a velocity lower by a factor of five compared to its lighter counterpart, while keeping the wind power approximately the same. In contrast, the high-power wind (W44) has a density similar to the light W43 wind but a velocity that is roughly twice as high. Thus, both the denser low-power wind (W43-dense) and the high-power wind (W44) carry a momentum flux roughly four times that of the light low-power wind (W43-light). However, the mass flux injected differs significantly: it is about twenty times higher for the dense wind, but only twice as high for the high-power wind. This difference in mass loading has important consequences for the dynamics and stability of the outflows. The dense wind, with its high mass flux and slower velocity, propagates more steadily and remains relatively stable, whereas the lighter high-power wind, despite having a similar momentum flux, is more prone to recollimations, shear instabilities at the edge, and turbulent mixing. These aspects of propagation and stability in light versus dense winds have been discussed in detail in Paper~II.

Jet simulations were performed on a $512\times512\times512$ grid with dimensions $4~\kpc \times 4~\kpc \times 4~\kpc$, yielding a spatial resolution of $\sim7.8$~pc. With an injection region diameter of 100~pc, the jet width is resolved by 12 cells. The wind simulations use a wide domain of $8~\kpc \times 8~\kpc \times 4\kpc$ with $800\times800\times400$ cells and a resolution of 10~pc. A larger domain is used for the winds, as they are wide-angled and are expected to have wider cocoons when compared to the jets. The wind injection region has a diameter of 400~pc and is resolved by 40 cells. All of the above constitute the compact jet and wind runs. In addition, we have carried out a large-scale (length of 20~kpc) jet simulation with a 200~pc diameter and a power of $10^{45}~\ergs$, resolved at 15.6~pc. This corresponds to 12 cells across the injection region diameter. This run is denoted as J45 in Table~\ref{tab:sim_table}.

\subsection{CREs injection in simulations and spectral modeling}
\label{sec:cre-setup}
In this study, we use the  Lagrangian Particle module \citep{vaidya_2018} of the \textsc{pluto} code to study the evolution and non-thermal emission from CREs. This module models the CREs as an ensemble of microparticles (i.e., electrons) that follow the fluid in the Eulerian grid. The phase-space evolution of the CREs is performed by solving the relativistic Boltzmann equation \citep{webb_1989}, which also accounts for the acceleration caused by diffusive shock acceleration (DSA) and various radiative cooling processes, including energy loss due to adiabatic expansion, synchrotron radiation, and inverse Compton scattering of particles by the surrounding Cosmic Microwave Background (CMB) photons.
 
CREs are injected into the simulation domain in two primary regions: just above the jet or wind inlet along the $Z$-axis, and ahead of the forward shock. Along the jet/wind axis, two CRE macroparticles are injected per cell in all simulations, except in J45, where four particles per cell are used. The injection frequency is tailored for each simulation to ensure adequate filling of the cocoon\footnote{For jets, CREs are injected every 10th time step in J43, and at every time step in J44 and J45. For winds, particles are injected every 5th time step. For J43, convergence tests were performed until the jet reached a height of 1~kpc, beyond which the computational cost became prohibitive. Increasing the CRE injection frequency to every 5th time step did not significantly alter the resulting flux values, indicating that the results are converged as long as the cocoon remains sufficiently well filled.}. 

Additional CREs are injected ahead of the forward shock\footnote{CREs are injected every 200th time step in J43, and every 80th time step in the other simulations. No forward shock injection is performed for J45.}, where the pre- and post-shock regions are determined using pressure and density thresholds defined relative to the initial halo values, \(p_0\) and \(\rho_0\), as done in our previous papers. Different pressure thresholds are used for different simulations because the pressure jump at the forward shock can vary across different powers of jets and winds. Our earlier studies (Paper I and II) showed that compact jets and high-power winds produce stronger pressure jumps than low-power winds. These thresholds allow us to identify the high-pressure shocked region and inject CREs in the corresponding pre-shock cells. Thus, we adopt pressure thresholds of $4p_0$ for jets, $2p_0$ for W43-light, and $1.5p_0$ for the W43-dense, while a threshold of \(4p_0\) is used for the W44-light. In the W43 runs, an additional density threshold of \(1.5\rho_0\) is applied to more accurately identify regions suitable for CRE injection ahead of the weakening forward shock at later times, particularly in the lateral directions. Finally, to exclude material belonging to the jet cocoon from the above analysis, we impose an additional constraint requiring the jet tracer\footnote{In \textsc{pluto}, tracers are scalar quantities passively advected with the fluid. Initially, the jet tracer (tr1) is set to 1 in the injection region and 0 elsewhere.} to satisfy \({\rm tr1} < 10^{-20}\).

After we detect the forward shock region, one CRE is added every cell ahead of the shock front. Across all simulations, several million CREs are typically injected, as listed in the last column of Table~\ref{tab:sim_table}.

At injection, the CREs are initialized with a steep power-law spectrum and limiting Lorentz factors ($\gamma_{\rm min}, \gamma_{\rm max}$) = ($10^2, 10^6$). Their initial spectra, in terms of energy, are given as

\begin{align}
    N(E) = n_m \left(\frac{1-\delta}{E_{\rm max}^{1-\delta} - E_{\rm min}^{1-\delta}} \right)E^{-\delta}
\end{align}

where $\delta = 9$\footnote{CREs are initialized with a very steep power-law energy spectrum} and $n_m$ is the CREs number density, i.e. $n_m = \int_{E_{\rm min}}^{E_{\rm max}} N(E) dE$. Therefore, the value for $n_m$ is given as \citep[see also][]{mukherjee_2021a},
\begin{equation}
    n_m = f_E \frac{\Gamma}{\Gamma - 1} \frac{\delta -2}{\delta - 1}\frac{p_j}{n_c\gamma_{\rm min}m_e c^2}
\end{equation}
In the above equation, $p_j$ denotes the pressure of the jet, $f_E$ is the fraction of internal energy density carried by CREs, $\Gamma = 5/3$ is the adiabatic index for the ideal gas, and $n_c$ is the number of CRE particles injected in a computational cell.

To update the final spectrum of CREs after a shock, we adopt a convolution-based spectral update method, following the approach of \citet{mukherjee_2021a}. In several earlier studies \citep{mimica_2009,fromm_2016,vaidya_2018}, the spectrum of shock-accelerated particles was reset to a power-law with an index determined by diffusive shock acceleration (DSA) theory. However, this erases the spectral history accumulated from previous shock encounters. To preserve this history, we compute the post-shock (downstream) spectrum, ($N_{\rm dn}(E)$), by convolving the upstream spectrum, ($N_{\rm up}(E)$), with the DSA-predicted spectrum, ($G_{\rm DSA}$). The updated CRE spectrum is therefore given by:

\begin{equation}
    N_{\rm dn} (E) = C \int_{E_{\rm min}}^{E_{\rm max}} N_{\rm up}(E')\, G_{\rm DSA} (E,E') \,\frac{dE'}{E'}
    \label{eq:upstream_spectra}
\end{equation}

where $G_{\rm DSA} (E,E') = (E/E')^{-q+2}$, for $E \in (E_{\rm min}, E_{\rm max})$. The minimum energy ($E_{\rm min}$) is set equal to the minimum energy of the upstream spectrum. The maximum energy ($E_{\rm max}$) is estimated by equating the synchrotron cooling timescale with the particle acceleration timescale, following \citet{mimica_2012,vaidya_2018}. The spectral index ($q$) is determined based on the orientation of the shock with respect to the magnetic field - parallel or perpendicular - using results from semi-analytical studies of DSA in relativistic shocks \citep{keshet_2005, takamoto_2015}, as adopted in \citet{vaidya_2018}.

The compression ratio may exceed 4 in some shocks because of numerical errors, and we found this to be the case for less than 5 percent of the shocked CREs. For such CREs, the particle spectra are updated by assuming an asymptotic spectral index of $q=4.23$, consistent with the ultra-relativistic limit predicted by \citet{kirk_2000}. A computational cell is considered shocked for the purposes of DSA if the relative thermal pressure gradient is greater than 3, thereby ensuring the selection of strong shocks.

The normalization constant $C$ is determined in two steps to ensure that the local maximum energy and number density of the CREs are within specified fractions of the fluid, as done in \citet{mukherjee_2021a}. The normalization is first set to preserve the energy densities of the particles to be a fraction ($f_\varepsilon$) of the fluid internal energy density ($\varepsilon =  p/(\Gamma - 1)$). Additionally, if the chosen normalization results in the CREs' number density becoming higher than a fraction ($f_N$) of the fluid number density, the previously obtained normalization is lowered to ensure an exact match. In that case, the total CREs energy density will be lower than the threshold of $f_\varepsilon \times \varepsilon$, due to a lower value of the new normalization. Thus, the above two-step procedure ensures that the CREs have energies and number densities that are within the prescribed fractions $f_\varepsilon$ and $f_N$ for energy and mass, respectively. For this work, we have assumed $f_\varepsilon = 0.1$ everywhere, and $f_N$ inside the cocoon is equal to the local jet-tracer value and is 0.1 for the shocked ambient medium (SAM) and the forward shock (where jet-tracer is close to zero). In the following sections, we examine the evolution of electrons injected into the jets and winds and variations in multi-frequency synchrotron emission and spectral indices.

\section{Results}
\label{sec:results}
\subsection{Comparing CREs evolution in jets and winds}
\label{sec:cre_evolve_jw}
In this section, we focus on the evolution and dynamics of CREs in the compact jet and wind simulations performed in this study.
 The slices in the $Y$-$Z$ plane depicting the maximum Lorentz factor attained by CREs are shown in Fig.~\ref{fig:lorentz_jw}. The slice widths are widened for the J44 and W43-light to have sufficient filling in their cocoons as their counterparts. Fig.~\ref{fig:shkn_jw} illustrates trajectories of some selected CREs launched along the jets and winds, and the probability distribution functions (PDFs) of the number of shocks encountered by CREs in different simulations are shown in Fig.~\ref{fig:pdf_shkn}. Key differences in the evolution of particles in the jets and winds, as inferred from the above figures, are outlined below:

\begin{itemize}
\item \textbf{Distribution of maximum Lorentz factors}\\
\underline{\textit{Jets}}:\, In both jets (J43 and J44), the CREs with the highest Lorentz factors ($\gamma_{\rm max} \gtrsim 10^{7}$) are found mainly along the jet spine and near the head (see Fig.~\ref{fig:lorentz_jw}). These values are larger than the maximum injected energies of $10^6$ (Lorentz factors) during the injection. Those with intermediate $\gamma_{\rm max}$ values ($10^{5}$–$10^{7}$) are distributed throughout the central regions of the cocoon, with notably lower values in the lower-$Z$ regions, e.g., see J44. In the SAM, the high $\gamma_{\rm max}$ are seen near the forward shock.\\

\underline{\textit{Winds}}:\, 
In winds, the CREs are energized to high $\gamma_{\rm max}$ values upon crossing the shock at the Mach disc (see Fig.~\ref{fig:lorentz_jw}), and retain most of their energies while being advected along the lateral streams. Further downstream into the cocoon, $\gamma_{\rm max}$ gradually decreases. This trend is observed in both low and high-density wind cases. In the SAM, high $\gamma_{\rm max}$ values can be seen near the forward shock, which is more clearly visible for the light winds (e.g., see W44-light). In the dense wind, forward propagation is slower, causing the forward shock to broaden and weaken by the time the wind reaches $\sim4~\kpc$. As a result, several CREs at the later stages are not significantly accelerated at the forward shock in this case\footnote{We detect only strong shocks; hence, weaker forward shocks, particularly at later stages in the wind evolution, are not registered for CREs shock encounters.}.\\

\item \textbf{Variation in the number of shock encounters}\\
\underline{\textit{Jets}}:\, 
The left panel of Fig.~\ref{fig:shkn_jw} shows sample trajectories of CREs in jets. It is evident that these particles undergo multiple shock crossings, being repeatedly accelerated as they propagate along the jet spine toward the hotspot, before being redirected by the backflow. Some CREs experience more than six shock encounters before reaching the hotspot, and additional shocks are encountered near the hotspot region within the backflowing plasma from the jet head. Beyond this point, the particles follow the backflow into the cocoon, where they encounter fewer shocks. Nevertheless, complex internal shock structures within the cocoon may still contribute to re-acceleration, depending on the cocoon’s properties and the jet's variability \citep[see e.g.][]{mukherjee_2021a,dubey_2023}.\\

\underline{\textit{Winds}}:\, The sample trajectories of CREs in winds are shown in Fig.~\ref{fig:shkn_jw}, with the middle panel representing the light wind and the dense wind in the right panel. In these cases, most CREs undergo primary shock acceleration at the Mach disc, while turbulent motions and vortices within the cocoon occasionally generate additional shocks. The vortices are more pronounced in the light wind (middle) than in the dense wind (right), primarily due to stronger backflows in the former.  

The distribution of shock encounters for CREs in jets and winds, shown in Fig.~\ref{fig:pdf_shkn}, further highlights that CREs in jets undergo more shock crossings compared to those in winds.\\

\end{itemize}

\subsection{Joint CRE spectra from jets and winds}
\label{sec:join_spec_jw}

\begin{figure*}
 \centerline{
\def\arraystretch{1.0}
\setlength{\tabcolsep}{0.0pt}
\begin{tabular}{lcr}
      \includegraphics[width=\linewidth,keepaspectratio]{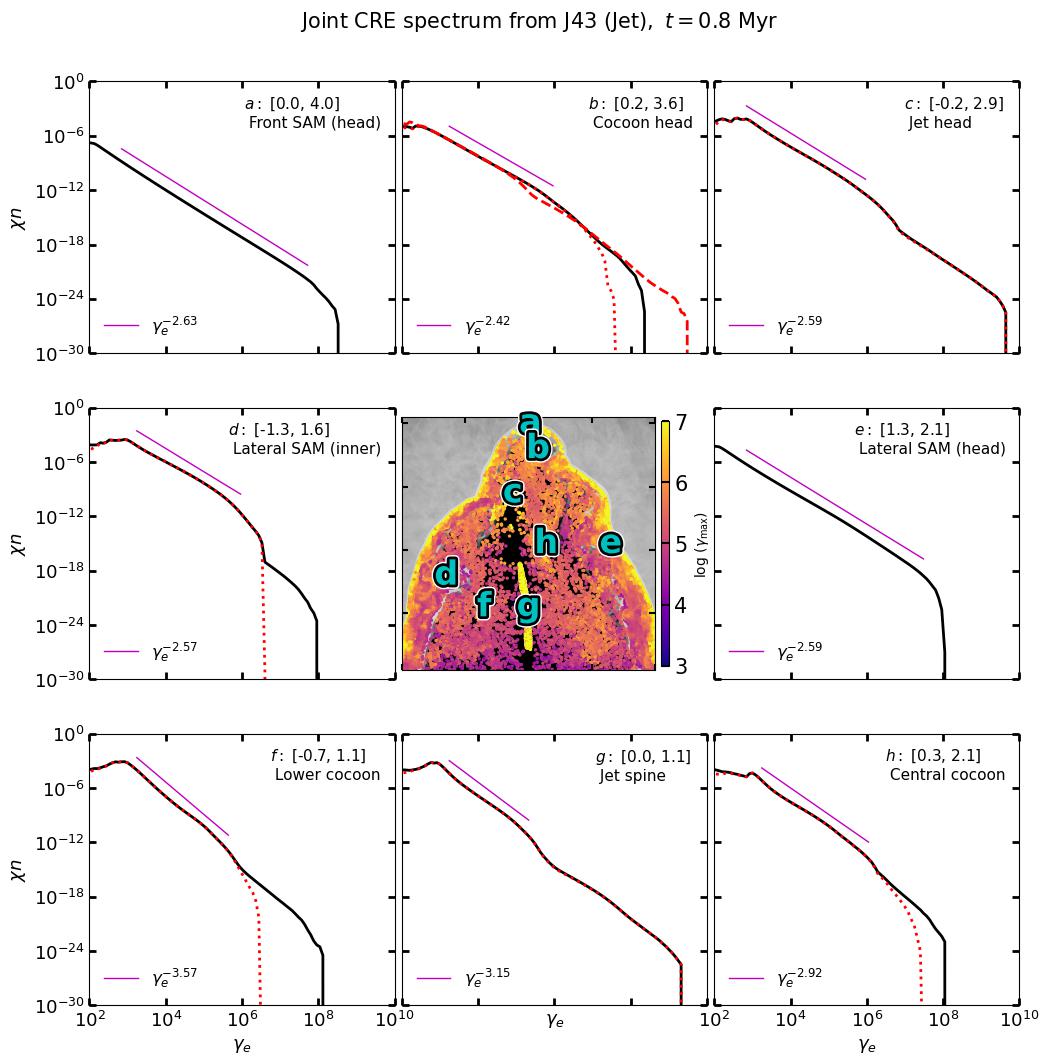} &
     \end{tabular}}
    
\caption{LOS-integrated joint CREs spectrum for selected domains of size $200~{\rm pc}^2$ on the $Y$--$Z$ image plane for J43 at 0.8~Myr. Here $\chi n$ represents the number density of CREs for the given value of the Lorentz factor $\gamma_e$. Coordinates for the midpoints of the domain are listed in the upper right corner of each plot. In each subplot, we indicate the location on the middle panel along the LOS; however, the LOS also traverses the cocoon and SAM, whose emission contributes to the total spectra shown. The dotted spectra indicate the spectrum obtained solely from the cocoon, excluding SAM. Same as Fig.~\ref{fig:lorentz_jw}, the background displays the logarithmic density, with darker areas indicating lower gas density.}
 \label{fig:43j_joinspec}
\end{figure*}

\begin{figure*}
 \centerline{
\def\arraystretch{1.0}
\setlength{\tabcolsep}{0.0pt}
\begin{tabular}{lcr}
      \includegraphics[width=\linewidth,keepaspectratio]{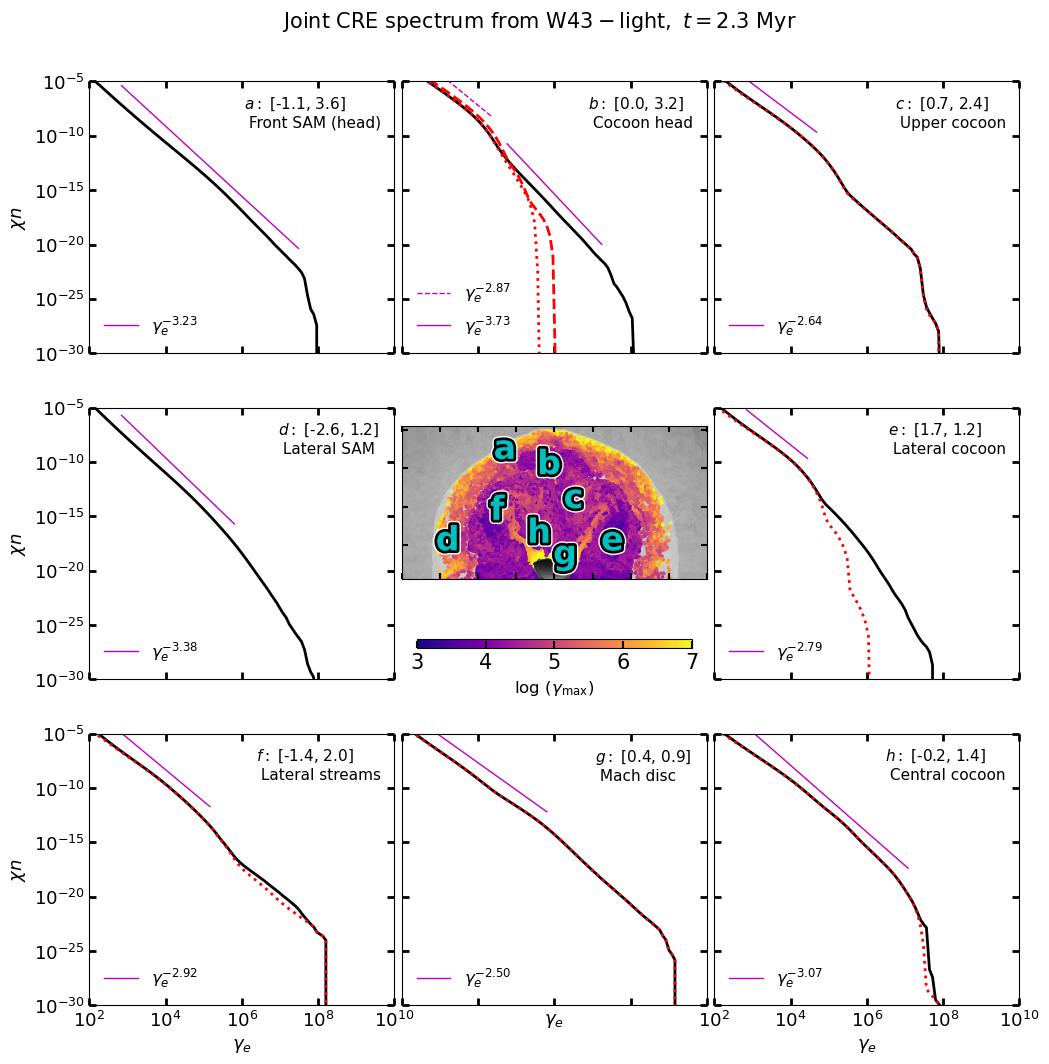} &
     \end{tabular}}
    
\caption{LOS-integrated joint CREs spectrum for selected domains of size $500~{\rm pc}^2$ on the $Y-Z$ image plane for W43-light at 2.3~Myr. Coordinates for the midpoints of the domain are listed in the upper right corner of each plot. In each subplot, we indicate the location on the middle panel along the LOS. The dotted spectra indicate the spectrum obtained solely from the cocoon. Same as Fig.~\ref{fig:lorentz_jw}, the background displays the logarithmic density, with darker areas indicating lower gas density.}
 \label{fig:43wl_joinspec}
\end{figure*}

\begin{figure*}
 \centerline{
\def\arraystretch{1.0}
\setlength{\tabcolsep}{0.0pt}
\begin{tabular}{lcr}
      \includegraphics[width=\linewidth,keepaspectratio]{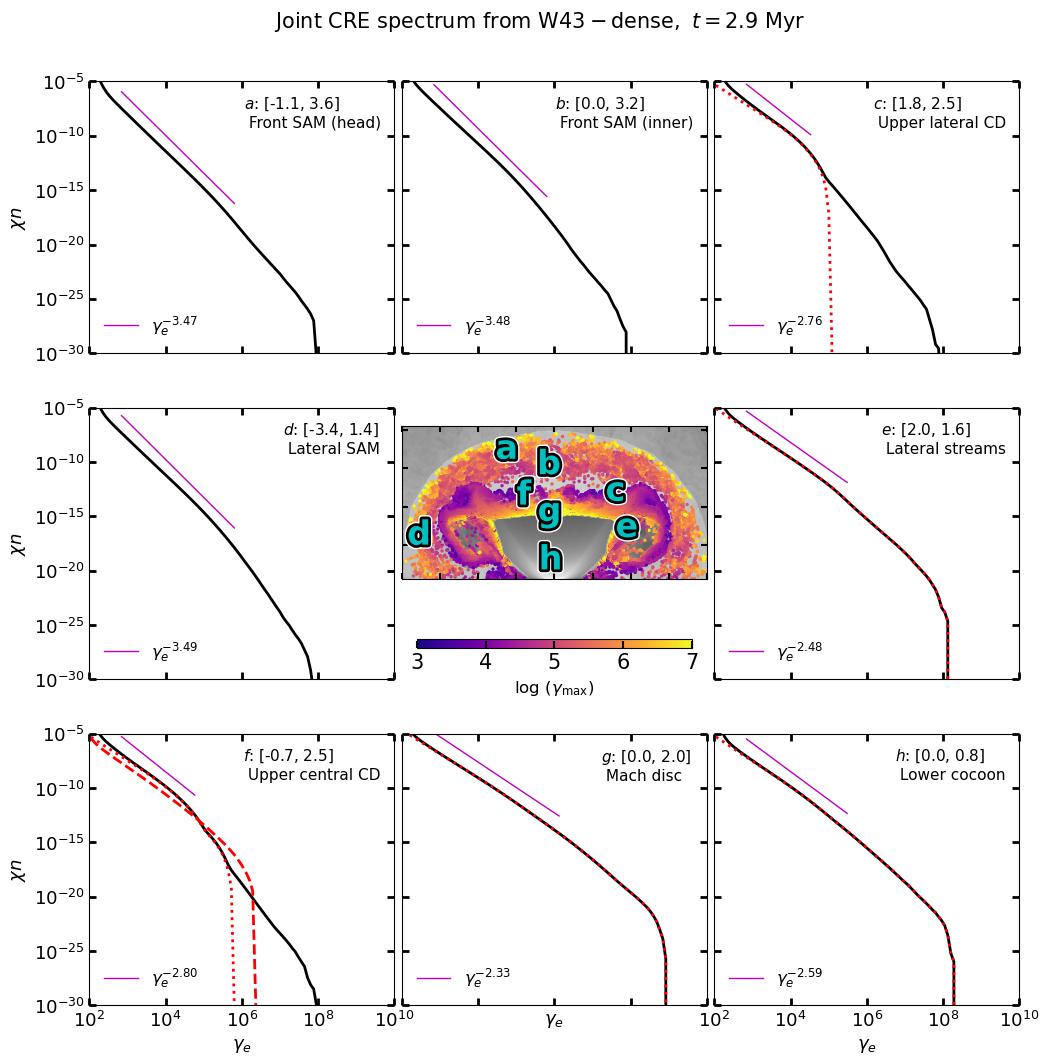} &
     \end{tabular}}
    
\caption{Same as Fig.~\ref{fig:43wl_joinspec} for W43-dense at 2.9~Myr. CD denotes the contact discontinuity in the figure above. Same as Fig.~\ref{fig:lorentz_jw}, the background displays the logarithmic density, with darker areas indicating lower gas density.}
 \label{fig:43wh_joinspec}
\end{figure*}

\begin{figure*}
 \centerline{
\def\arraystretch{1.0}
\setlength{\tabcolsep}{0.0pt}
\begin{tabular}{lcr}
      \includegraphics[width=0.99\linewidth,keepaspectratio]{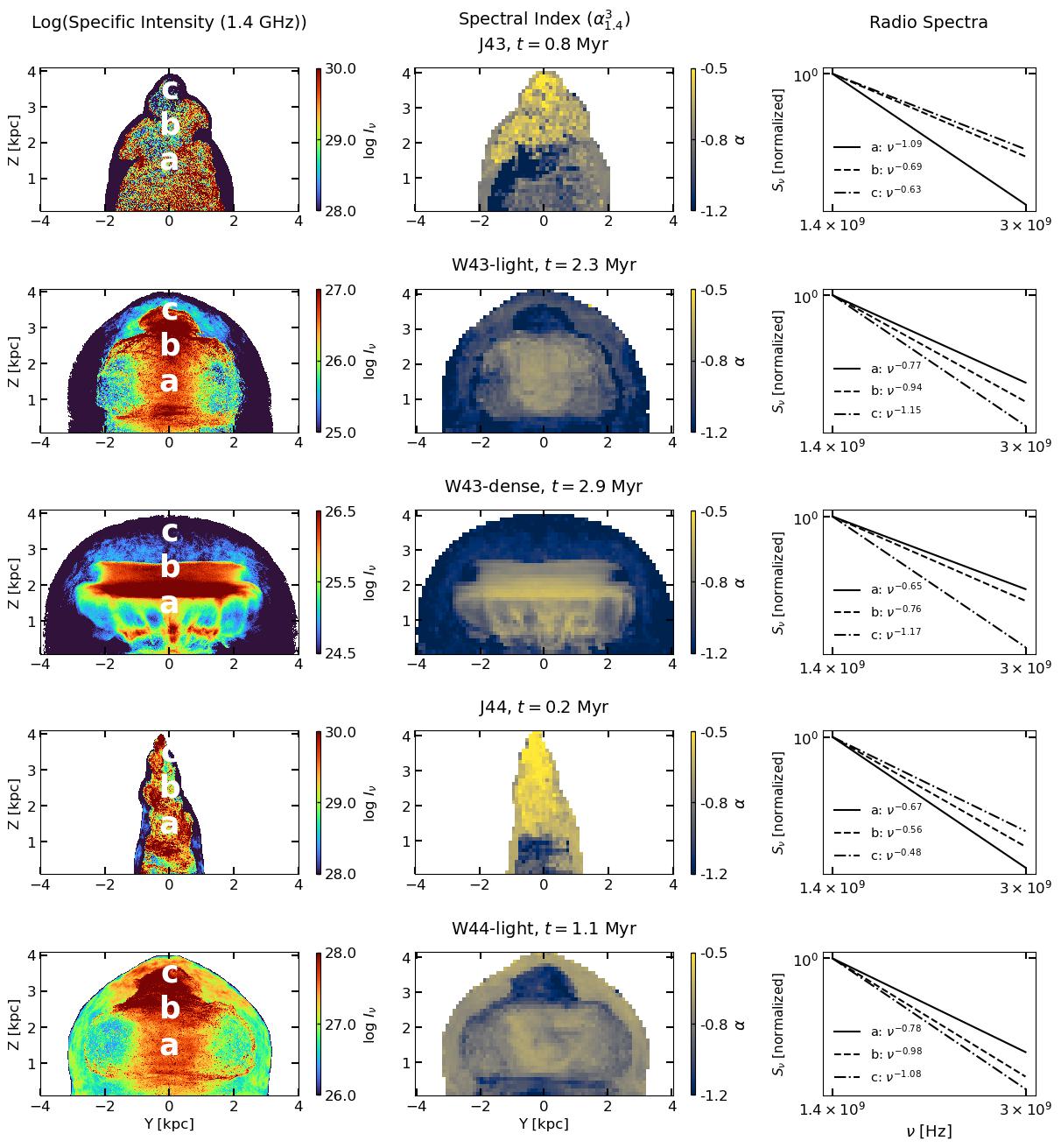}
     \end{tabular}}
\caption{Left to right: Logarithmic synchrotron flux ($\log I_{\nu}$ [$\flux$]), at 1.4 GHz; Spectral index maps for 1.4-3 GHz; Spatial radio spectra for $1\kpc \times 1~\kpc$ regions $a, b,$ and $c$ (indicated in the left panels), for both jets and winds. All panels are shown in the $Y$–$Z$ image plane ($\theta_I = 90^\circ$). A larger domain, similar to that used for the winds, is shown for the jets to maintain consistency in the collage.}
 \label{fig:synch_90deg_jw}
\end{figure*}

\begin{figure*}
 \centerline{
\def\arraystretch{1.0}
\setlength{\tabcolsep}{0.0pt}
\begin{tabular}{lcr}
      \includegraphics[width=0.95\linewidth,keepaspectratio]{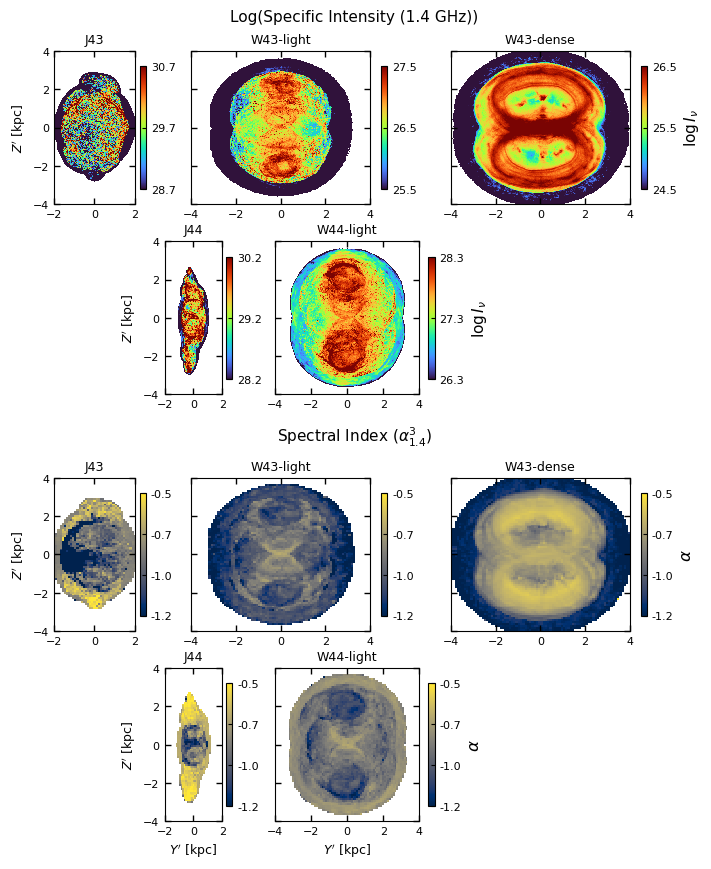} &
     \end{tabular}}
    
\caption{Logarithmic specific intensity ($\log I_{\nu} [\flux]$) at 1.4~GHz, and spectral index maps at  $45^\circ$ LOS for 1.4-3~GHz. The indices are computed over $100~{\rm pc}^2$ regions on the image plane.}
 \label{fig:spec_45deg_jw}
\end{figure*}

\begin{figure*}
 \centerline{
\def\arraystretch{1.0}
\setlength{\tabcolsep}{0.0pt}
\begin{tabular}{lcr}
      \includegraphics[width=0.85\linewidth,keepaspectratio]{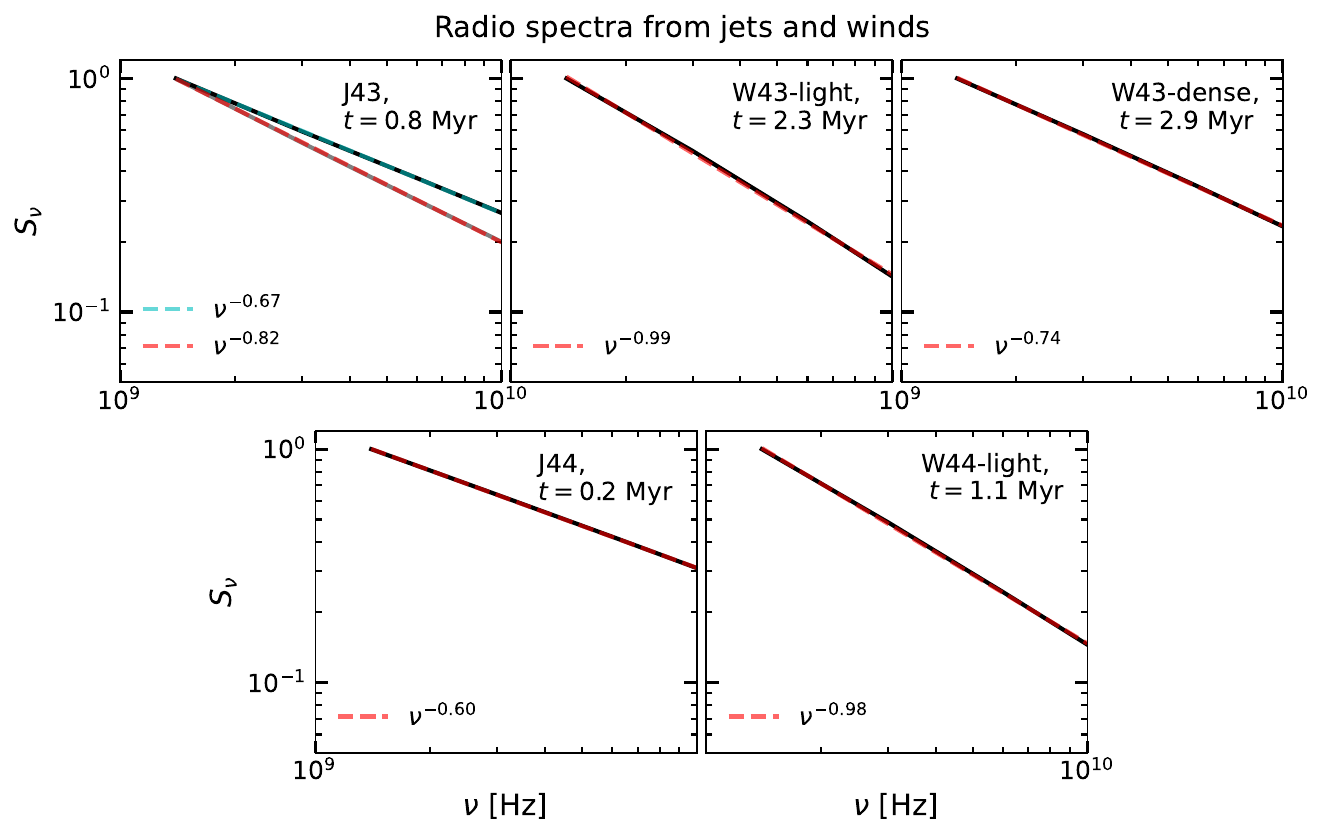} &
     \end{tabular}}
    
\caption{Flux-integrated radio SEDs from the compact jet and wind simulations presented in this study. For each model, the SED is fitted with a power law over 1.4-10~GHz, depicted using a dashed red line. For J43 (top left), an additional SED is extracted from a region near the jet head ($Z\gtrsim 2~$kpc); the corresponding fits are shown with a dashed cyan line.} 
 \label{fig:tot_spectra}
\end{figure*}

\begin{table*}
	\centering
	\caption{Flux- and volume-weighted mean spectral indices for jets and winds, estimated from Fig.~\ref{fig:synch_90deg_jw}. The values are shown over the $1.4$–$3~\mathrm{GHz}$ and $0.15-3~$GHz frequency ranges.}
	\label{tab:index_table_2}
	
	\begin{tabular}{|c|c|c|c|c|c|c|} 
		\hline
Simulation  &   \multicolumn{2}{|c|}{Cocoon ($1.4-3$~GHz)}  & \multicolumn{2}{|c|}{Cocoon + SAM} ($1.4-3$~GHz)  & \multicolumn{2}{|c|}{Cocoon ($0.15-3$~GHz)} \\ \hline
 &   Flux wtd.  & Volume wtd. &  Flux wtd.  & Volume wtd.  &  Flux wtd.  & Volume wtd. \\
		\hline
			$\mathrm{J43} \,$ & -0.85 & -0.784 & -0.85 & -0.784  & -0.785 & -0.79 \\
	\hline
    $\mathrm{J44}\,$ & -0.59 & -0.61 & -0.58 & -0.62 & -0.55  & -0.59 \\
			\hline
	$\mathrm{W43-light} \,$ & -0.94 & -1.03 & -0.94 & -1.03 & -0.83 & -0.87\\ 
		\hline
           $\mathrm{W44-light}\, $& -0.97 & -0.95 & -0.94 & -0.86  & -0.87 & -0.9  \\
		\hline
	$\mathrm{W43-dense}\, $ & -0.71 & -0.9 & -0.72 & -1.0 & -0.68 & -0.84\\
			\hline       

	\end{tabular}

\end{table*}

The spectrum of an individual CRE macro-particle evolves due to losses (adiabatic and radiative) and shock acceleration. However, on an image plane, the joint spectrum is a combination of the individual spectra originating from different macro-particles along the line of sight. Therefore, we have computed the LOS-integrated joint spectra of some domains of size $200~{\rm pc}^2$ for J43 and $500~{\rm pc}^2$ for the winds on the $Y-Z$ image plane. Different domain sizes were selected to ensure adequate sampling of the CREs: a larger domain for winds, reflecting their broader cocoons, and a correspondingly smaller one for jets. The resulting spectra are displayed in Figs.~\ref{fig:43j_joinspec} to~\ref{fig:43wh_joinspec}, which are computed at times similar to that in Fig.~\ref{fig:lorentz_jw}, i.e., when the jet and winds reach a height of 4~kpc. These spectra are complex, but the low-energy region ($\gamma_e\lesssim 10^6$) mostly behaves as a power-law. Thus, we fit the joint spectra at low energies using a power-law. The fit is represented by the magenta line. These spectra include contributions from the cocoon as well as the SAM along the LOS. Thus, we have also plotted the joint spectra from the cocoon alone using dotted red curves. The spectra originating from regions near the head of the jet/wind outflow at earlier times, when they had reached a height of approximately (2~kpc), are also shown as red dashed curves. The central panel shows a log-scaled color map of the maximum Lorentz factor distribution (in log), with overlaid markers ($a-h$) indicating regions where the spectra in the surrounding panels were extracted. The behavior of these spectra is discussed below:

\begin{itemize}
    \item \textbf{J43:}\, 
The joint spectra for the jet (Fig.~\ref{fig:43j_joinspec}) exhibit noticeable variations across different locations, both in the spectral slope and in the range of electron energies ($\gamma_e$). The spectra extend to the highest energies ($\gamma_e \gtrsim 10^9$), particularly when coinciding with regions along the jet spine ($g$) and near the head ($c$). Comparing spectra from the upper cocoon regions ($b$: $\gamma_e^{-2.42}$, $c$: $\gamma_e^{-2.59}$) with those from the lower regions ($f$: $\gamma_e^{-3.57}$, $h$: $\gamma_e^{-2.92}$) reveals a steepening of the CRE spectra, as they propagate from the jet head downward into the cocoon. The dashed red curve (in $b$) shows that even at earlier times, when the jet was only $\sim2$~kpc in size, the spectrum near the cocoon head exhibited a slope similar to that at later times. However, the spectrum extended to substantially higher energies at these earlier stages than at later times, suggesting that radiative cooling progressively dominates the high-energy CRE population, even near the cocoon's head.

In locations where the total spectra include contributions from both the cocoon and the SAM (except $a$ and $e$), the flatter part of the spectrum at $\gamma_e \lesssim 10^6$ is primarily contributed to by the cocoon, as indicated by the dotted red curves. From regions of frequent CRE acceleration ($c$ and $g$), the CREs in the cocoon dominate the total spectrum. The spectra from the SAM extend to higher $\gamma_e$ values (up to $\sim10^8$) than those from the lower cocoon regions ($f$ and $h$), owing to the longer synchrotron cooling times of CREs in the SAM.

\item \textbf{W43-light:}\, 
In light winds (Fig.~\ref{fig:43wl_joinspec}), the flattest spectrum - modeled at $\gamma_e \lesssim 10^6$ - originates from regions just ahead of the Mach disc (e.g., $g$: $\gamma_e^{-2.5}$). The spectra become progressively steeper further downstream (e.g., $c$, $f$, and $h$). This steepening occurs because, after crossing the shock at the Mach disc, the CREs are advected along lateral streams (Paper~II) and subsequently carried into the backflowing plasma. During this journey, they encounter a few additional shocks (see Fig.~\ref{fig:shkn_jw}) and primarily experience radiative losses within the cocoon. The spectra through the cocoon head (in $b$) are broadly similar at earlier times (red dashed curves) and later times (red dotted curves), although the former extend to somewhat higher energies. This suggests that these regions in the cocoon of the light-wind case are predominantly filled with cooling CREs.  

In the SAM, the spectrum at the head is notably steeper ($a: \sim \gamma_e^{-3.2}$) than in the jet case ($ a: \sim \gamma_e^{-2.6}$). This suggests that, \textit{by the time both structures reach comparable spatial extents, the forward shock driven by winds can get weaker than that produced by a jet of similar power.} The low-energy part of the joint spectra is primarily dominated by cocoon (except in $a$ and $d$), and is characterized by a flatter slope. In contrast, the spectra from the SAM ($a$ and $d$) extend to higher $\gamma_e$ values and exhibit steeper slopes.

\item \textbf{W43-dense:}\, 
In the dense wind, the spectra distribution (see Fig.~\ref{fig:43wh_joinspec}) shows several similarities with the light wind case in Fig.~\ref{fig:43wl_joinspec}.  The flattest CRE spectrum originates from regions just ahead of the Mach disc (e.g., $g$: $\gamma_e^{-2.33}$), similar to what is observed in W43-light. Also, as one moves downstream from the Mach disc to high $Z-$regions, the spectra become progressively steeper (up to $\gamma_e^{-2.8}$ in regions $c$ and $f$) and are limited to lower maximum energies as compared to region $g$. However, shocks within the backflows or vortices can also accelerate CREs to high energies, as seen in regions $e$ and $h$ within the cocoon. This additional acceleration keeps the spectra from these regions extended to higher energies ($\gamma_e \gtrsim 10^8$) and less steep ($\gamma_e^{-2.48}$). The spectra through the wind's cocoon head (i.e. in panel $f$) are shown at earlier times with red dashed curves. The slope remains similar to as that oroginating at later stages (red dotted curves), but extends to a bit higher energies. Similar to the light wind in Fig.~\ref{fig:43wl_joinspec}, the spectra through the cocoon head (panel $f$) at earlier times (red dashed curves) exhibit a slope very similar to that at later stages (red dotted curves). However, the former spectra extend to somewhat higher energies, indicating radiative cooling for the high-energy CRE population as the wind evolves.

The spectra obtained from the front head of the SAM here are a bit steeper ($a: \gamma_e^{-3.47}$) when compared to the light wind ($a: \gamma_e^{-3.2}$), indicating a comparatively weak forward shock. Interestingly, the slopes in both cases are steeper than the jet, \textit{indicating a weaker forward shock strength in the wind cases compared to the jet.}

\end{itemize}

\subsection{Multi-frequency emission from jets and winds}
\label{sec:synch_jw}

In this section, we examine the synchrotron emission and spectral characteristics of compact jets and winds discussed in previous sections. The corresponding analysis is presented in Figs.~\ref{fig:synch_90deg_jw}-\ref{fig:tot_spectra} and discussed in detail below.

\subsubsection{Morphology of synchrotron emission}
\label{sec:synch_morphlogy}
The specific intensity maps for compact jets and winds, integrated onto an image plane oriented at $\theta_I = 90^\circ$, are shown in Fig.~\ref{fig:synch_90deg_jw} (left panels). These maps show the fluxes at a radio frequency of 1.4~GHz. Since the emitted flux decreases with increasing frequency, the colorbar extent is chosen to better emphasize the general emission morphology and highlight the brightest regions. The emission maps for a higher radio frequency value, i.e., 3~GHz, are shown in Fig.~\ref{fig:synch_90deg_jw_hf} in the Appendix. Additionally, emission maps at an inclined LOS of $\theta_I = 45^\circ$ are displayed in Fig.~\ref{fig:spec_45deg_jw}. Since the simulation domain covers only the upper half, the integrated maps at inclined viewing angles are generated by first estimating the emissivities at $45^\circ$ (upper domain) and $135^\circ$ (corresponding to the lower domain) separately. A new emissivity data cube is then constructed by joining the lower and the upper domains. This combined cube is subsequently integrated along the observer’s line of sight to obtain the flux distribution for the inclined view.

Below, we discuss the spatial flux variations across the jets and winds.

\underline{\textit{Emission from jets:}}\, 
The specific intensity maps at an LOS of $90^\circ$ for J43 and J44 at radio frequencies are shown in the left panels of Figs.~\ref{fig:synch_90deg_jw} and~\ref{fig:synch_90deg_jw_hf}. The cocoon is prominently bright, while typical jet features such as a distinct spine (in both) and a hotspot (in J43), are not clearly visible at either frequency. This behavior contrasts sharply with the results presented in Paper~I, where the spine and hotspot appeared as the brightest regions in the $90^\circ$ maps (e.g. see Fig.~9 in Paper~I). Those earlier maps were generated using a fixed CRE spectral index, causing the emission to mainly follow the distributions of pressure and magnetic field. Consequently, regions of highest pressure and magnetic field - namely, the hotspot and spine - stood out as the brightest. In the present analysis, where the spectral evolution of CREs is included, the emission morphology differs markedly. At early times, however, the jets displayed a bright and well-defined spine and hotspot, before the backflowing CREs had filled the entire cocoon. As the system evolved and the cocoon became saturated with backflowing particles, these distinct features faded. 

Hence, in J43, the cocoon is uniformly filled with accelerated CREs due to frequent kinks\footnote{current-driven instability \citep{bodo_2013,bromberg_2016}, see review \citet{mukherjee_2025} for details} and changes in the jet direction, which mix the particles from earlier hotspot positions throughout the cocoon. As a result, the central cocoon becomes prominently bright, while the hotspot and its vicinity appear comparatively faint.

In contrast, the J44 undergoes only a few kinks, resulting in a prominent hotspot with bright, filament-like features emerging from it. These features can also be seen in the maps at an inclined LOS in Fig.~\ref{fig:spec_45deg_jw}. Unlike in J43, the regions closer to the jet head remain bright in this case. In both jets, the SAM shows substantially weaker flux than the cocoon, largely due to the shorter integration length along the line of sight through these regions.

\underline{\textit{Emission from winds:}}\,
Emission maps from the winds at 1.4~GHz are shown in Fig.~\ref{fig:synch_90deg_jw} for $\theta_I=90^\circ$, and in Fig.~\ref{fig:spec_45deg_jw} for $\theta_I=45^\circ$. The emission morphology closely resembles that seen in Paper~II. In light winds, a bright column-like feature is seen, originating from the regions between the lateral streams (same as Fig.~7 in Paper~II). 
At 1.4~GHz, the Mach disc is not distinctly visible but starts to become clearly identifiable in the 3~GHz maps, shown in Fig.~\ref{fig:synch_90deg_jw_hf}. The highest flux originates from bright horizontal arcs near the cocoon head, and also from the regions above the Mach disc, coinciding with lateral streams. These bright arcs result from CREs accumulating at the upper edges of the lateral streams. Thus, these arcs manifest as bright rings near the cocoon's head when viewed at inclined LOS in Fig.~\ref{fig:spec_45deg_jw}, which is similar to the findings of Paper~II (Fig.~11) for the light winds. Additionally, many of the older CREs are pushed into a hump-shaped region as new CREs emerge from the lateral streams, leading to enhanced low-frequency emission at $\theta_I=90^\circ$.

In the high-density wind case (W43-dense), the Mach disc and the region immediately ahead of it are the brightest features at 1.4 GHz, while the surrounding cocoon and the vortices show comparatively weak emission. As one goes to higher frequency (Fig.~\ref{fig:synch_90deg_jw_hf}), the emission becomes strongly concentrated around the Mach disc. For an inclined LOS (Fig.~\ref{fig:spec_45deg_jw}), this structure is projected as bright circular arcs or rings. Similar to the jet case, the flux from the SAM is significantly weaker than that from the cocoon.

\subsubsection{Spatial distribution of radio spectral indices}
\label{sec:spec_index}
The radio spectral index maps for the compact jet and wind runs are presented in the middle column of Fig.~\ref{fig:synch_90deg_jw} for $1.4-3$~GHz at an LOS of $\theta_I=90^\circ$. The indices ($\alpha$) are computed using flux measurements at the corresponding frequencies, using the expression:
$$\alpha_{\nu_1}^{\nu_2} = \log (S_2/S_1)/ \log (\nu_2/\nu_1)$$
where $S_{1}$ and $S_2$ corresponds to the volume integrated fluxes at the frequencies $\nu_1$ (lower) and $\nu_2$ (higher), respectively.
. We also present spatially resolved radio SEDs in the third column, motivated by situations where full-resolution data are unavailable and only multi-wavelength measurements from limited regions can be obtained. Specifically, we select spatial domains of area $1\kpc \times 1\kpc$ for this analysis, and along the direction of the flow, i.e., $Z$-axis. These domains are labeled $a,b,$ and $c$ on the emission maps in the first column, and the individual spectra from them, and the resultant indices are plotted in the third and fourth column plots, respectively. Additionally, spectral index maps for an inclined LOS, i.e. $\theta_I=45^\circ$, are shown in Fig.~\ref{fig:spec_45deg_jw}. Below, we list the insights drawn from these maps:

\underline{\textit{Jets}}:\,
A consistent trend across all cases is the progressive steepening of the spectrum from the tip of the head of the jet towards the lower regions of the cocoon, observable across both frequency ranges (Figs.~\ref{fig:synch_90deg_jw},~\ref{fig:spec_45deg_jw} and~\ref{fig:synch_90deg_jw_hf}). The flattest spectra appear at the jet head in both cases (with slopes of approximately $-0.6 ~{\rm to }-0.5$), becoming steeper further towards the base. This behavior supports the interpretation, as discussed in Sec.~\ref{sec:join_spec_jw}, that \textit{the most significant acceleration of CREs occurs at the jet's head, after which radiative cooling becomes dominant within the cocoon}. The indices from the $1\kpc^2$ domain in the right column for J43 exhibit the same trend discussed above: the spectra are relatively shallow near the jet head ($c:$ $\alpha\gtrsim$ -0.6) and progressively steepen toward the jet base ($a:$ $\alpha\approx$ -1.1). Similar features are seen in the maps for $\alpha_{1.4}^3$ produced at an inclined LOS in Fig.~\ref{fig:spec_45deg_jw}. Also, the trends and the values remain comparable till high frequencies, i.e. $\alpha_{3}^{10}$ in Fig.~\ref{fig:synch_90deg_jw_hf}. Similar features are also seen for J44, although the values are comparatively shallower than J43.

In jets, the SAM displays shallow spectral indices, with flatter values near the front shock head for both the jet runs. The forward shock in compact jets is strong as the SAM is narrow. However, it is important to note that the flux in these regions is often significantly lower than in the cocoon, potentially falling below detection thresholds. As a result, the spectral features from the SAM may not always be detectable.

\underline{\textit{Winds}}:\, The middle and right columns in Fig.~\ref{fig:synch_90deg_jw} show that in the light-wind cases, the spectra are shallowest at the Mach disc, with indices of around $\alpha \approx -0.75$. The indices show a steepening trend with moving to higher heights from the Mach disc within the cocoon (as seen from the last column in Fig.~\ref{fig:synch_90deg_jw}). As discussed in Sec.~\ref{sec:join_spec_jw}, \textit{in winds, CREs are primarily accelerated at the Mach disc and subsequently undergo radiative cooling}. This is more clearly seen at higher frequencies ($\alpha_{3}^{10}$ in Fig.~\ref{fig:synch_90deg_jw_hf}), where the spectral indices in the lateral cocoon and near the cocoon head become significantly steeper relative to their low-frequency values, indicating enhanced radiative aging. In contrast, the presence of freshly accelerated CREs just ahead of the Mach disc maintains relatively constant spectral indices. Similarly, flat indices near the Mach disc are also observed for inclined lines of sight (Fig.~\ref{fig:spec_45deg_jw}) at $1.4-3$~GHz, extending along the lateral streams and their heads. Notably, the regions between the lateral streams near the cocoon head display comparatively steep indices, which are not clearly seen in the $\theta_I=90^\circ$ maps (Fig.~\ref{fig:synch_90deg_jw}). This behavior arises from the rapid cooling of CREs after crossing the Mach disc, followed by advection through the lateral streams into the cocoon. When the emission is integrated along the line of sight through both the streams and the cocoon (for $\theta_I = 90^\circ$), these features become blended and difficult to separate, whereas they are clearly resolved for inclined viewing angles.

In W43-dense, for $1.4-3$~GHz, flatter spectral indices ($\sim-0.65$) are seen ahead of the broad Mach disc. The trend remains the same as the light wind here, and the spectra get steeper as one goes to heights from the Mach disc, and can be seen more clearly in the last two columns of Fig.~\ref{fig:synch_90deg_jw}. At higher radio frequencies (Fig.~\ref{fig:synch_90deg_jw_hf}), this trend becomes more apparent, and the flatter spectral indices have begun to be confined in regions close to the Mach disc (second figure in last row), suggesting cooling of high-energy CREs after the termination shock. 

In the SAM, the flatter values (although steeper than at the Mach discs) are seen mainly near the head in W43, whereas in W44, the shallow values are obtained from all parts of the SAM. This difference is attributed to narrow SAM in the high-power winds, and is similar to the trend seen for jets with increasing power. The colorbar indicates that these values are steeper when compared to their jet counterparts. The dense winds have a broad SAM, which has significantly reduced the strength of the forward shock's head, resulting in steep indices here.

\subsubsection{Unresolved radio SEDs and mean spectral indices}
In this section, we analyze the integrated properties of the compact jets and winds by constructing the corresponding radio SEDs (Fig.~\ref{fig:tot_spectra}), and computing the mean spectral indices across the entire domain (Table~\ref{tab:index_table_2}). Table~\ref{tab:index_table_2} lists the flux- and volume-weighted mean spectral indices averaged over the cocoon and cocoon+SAM regions for jets and winds shown in Fig.~\ref{fig:synch_90deg_jw}. We calculate the point-to-point spectral indices for observed frequencies from different radio surveys. We select 150 MHz (LoTSS:LOFAR Two-metre Sky Survey,
TGSS: TIFR GMRT Sky Survey), 1.4 GHz (FIRST: Faint Images of the Radio Sky at Twenty centimeters), and 3 GHz (VLASS: Very Large Array Sky Survey). These frequencies also represent a broad range that will be covered by the combination of the upcoming
Square Kilometer Array (SKA)-Low and SKA-Mid. The indices are estimated for ranges of $1.4-3$~GHz and $0.15-3$~GHz, using the fluxes estimated at these frequencies. The cocoon is identified using the jet-tracer thresholds, $\mathrm{tr1} >10^{-7}$ for jets and $\mathrm{tr1} >10^{-3}$ for winds, which are also used in our previous studies. To minimize contamination from aged CREs accumulating near the base due to reflecting boundaries, the averaging is restricted to regions sufficiently above the lower $Z$ boundary: $Z > 1$ for jets and $Z > 0.5$ for winds. For flux-weighted measurements, only regions with flux within 3 dex of the peak emission at 3~GHz are considered.

Similarly, regions above the core are selected to construct the integrated radio SEDs shown in Fig.~\ref{fig:tot_spectra}. The SEDs indicate that both jets and winds exhibit spectra that are well described by power-law behavior with almost the same index till 10~GHz frequency, as highlighted by the dashed red fits. The main results are summarized below:

\textbf{Jets:} In Fig.~\ref{fig:tot_spectra}, the low-power jet (J43) shows a steeper spectral index of $\alpha = -0.82$ for $1-10$~GHz range. However, when restricting the analysis to regions near the hotspot ($Z \gtrsim 2,\mathrm{kpc}$: dashed cyan fit), the spectrum becomes slightly flatter, with $\alpha \approx -0.67$, reflecting recently accelerated CREs. In contrast, the higher-power jet (J44) exhibits a shallower overall slope of $\alpha = -0.6$.

This behavior is consistent with the values listed in Table~\ref{tab:index_table_2}. Over the $1.4 - 3~\mathrm{GHz}$ range, J43 exhibits steep spectral indices ($\alpha \approx -0.85$) in flux-weighted and comparatively flat (-0.784) in volume-weighted measurements. The steepening in the former arises from the significant contribution of aged CREs residing in the central and lower parts of the cocoon, whereas the new CREs in the upper cocoon regions contribute weakly in comparison (see Fig.~\ref{fig:synch_90deg_jw}). The broad cocoon undergoes frequent mixing between pockets of reaccelerated CREs and older electron populations, driven by kink-induced disturbances, which collectively lead to a steeper integrated spectrum from the whole cocoon. In contrast, the more stable J44 jet gives a mean value of $-0.58$. When including the SAM in cocoon+SAM estimates, the trends and values remain largely unchanged due to the weak contribution of SAM to the total flux in both jets. The estimates remain almost similar at $0.15-3~$GHz, albeit some flattening is seen at the flux-weighted measurements.

\textbf{Winds:} In Fig.~\ref{fig:tot_spectra}, light winds (W43 and W44) produce consistently steep slopes of around $-1$ over the full frequency range, which is quite steeper than those obtained for the jets. Contrarily, the dense wind shows a slope of $-0.74$.

In Table~\ref{tab:index_table_2}, the flux-weighted mean values indicate that light winds tend to have steeper spectra than jets ($\alpha \approx -1.0 - -0.9$) for $1.4-3~$GHz, as also seen in Fig.~\ref{fig:tot_spectra}. Among the winds, the mean values from the dense winds ($-0.7$) are comparatively flat than light ones ($\approx -0.94$) with the same power (W43). This arises because, in dense winds, the bright regions at the Mach disc coincide with flatter spectral indices, while in light winds, the brightest flux occurs near the cocoon head where aged CREs accumulate. The flux and volume-weighted indices have comparable values in light winds. In contrast, in dense wind, the indices are steeper for volume-weighted than the flux-weighted, as the flux-weighted focuses on recently accelerated CREs near the Mach disc, but volume-weighted accounts for the cooling population in the cocoon, as can be inferred from the maps in Fig.~\ref{fig:synch_90deg_jw} for the W43-dense. When including the SAM in cocoon~+~SAM estimates, the trends and values remain largely unchanged due to the weak contribution of SAM to the total flux. A slightly flatter index is observed in the W44-light (volume-weighted), which is due to shallow values at the shock front (see W44-light in Fig.~\ref{fig:synch_90deg_jw}), when compared to the estimate from the cocoon only. 

At $0.15-3$~GHz, the inferred spectral indices for the winds appear narrower than those at higher frequencies. This suggests that certain particle populations in the winds are undergoing fast cooling at GHz frequencies, which is not that pronounced for the jet.

\begin{figure*}
    \centering
    \includegraphics[width=0.8\linewidth]{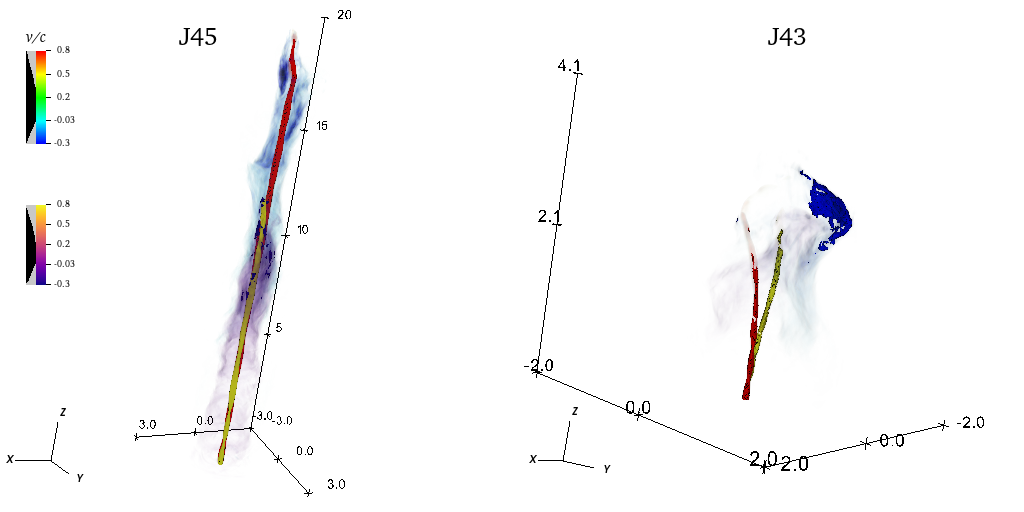}
    \caption{3D volume rendering of the $Z-$component of the velocity for the J45 (left) and J43 (right) jets. The plots are shown for two different times using different color-maps, i.e., 0.29~Myr (`plasma') and 0.41~Myr (`rainbow') for J45, and similarly 0.6~Myr and 0.8~Myr for J43, respectively.}
    \label{fig:visit_vz}
\end{figure*}

\begin{figure*}
\centering
\includegraphics[width=\linewidth, keepaspectratio]{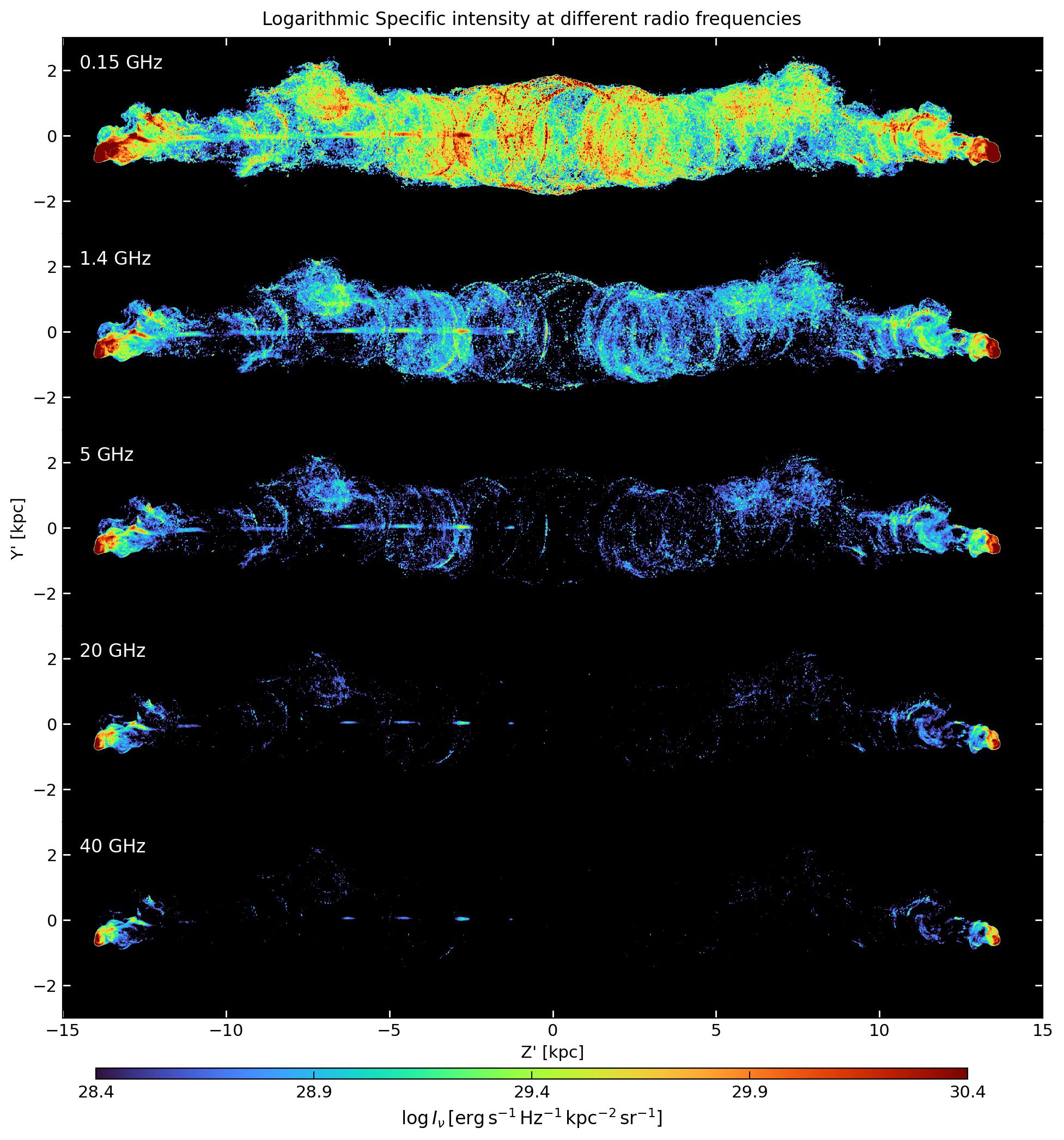}
\caption{Logarithmic surface brightness ($\log I_{\nu} [\flux]$) of J45 on an image plane oriented at $\theta_I = 45^\circ$. The emission is shown across radio frequencies from 0.15 to 40 GHz at 0.42 Myr. The colorbar range is fixed, and regions with values below the lower limit are excluded.}
 \label{fig:synch_45_45deg}
\end{figure*}

\begin{figure*}
\centering
\includegraphics[width=\linewidth, keepaspectratio]{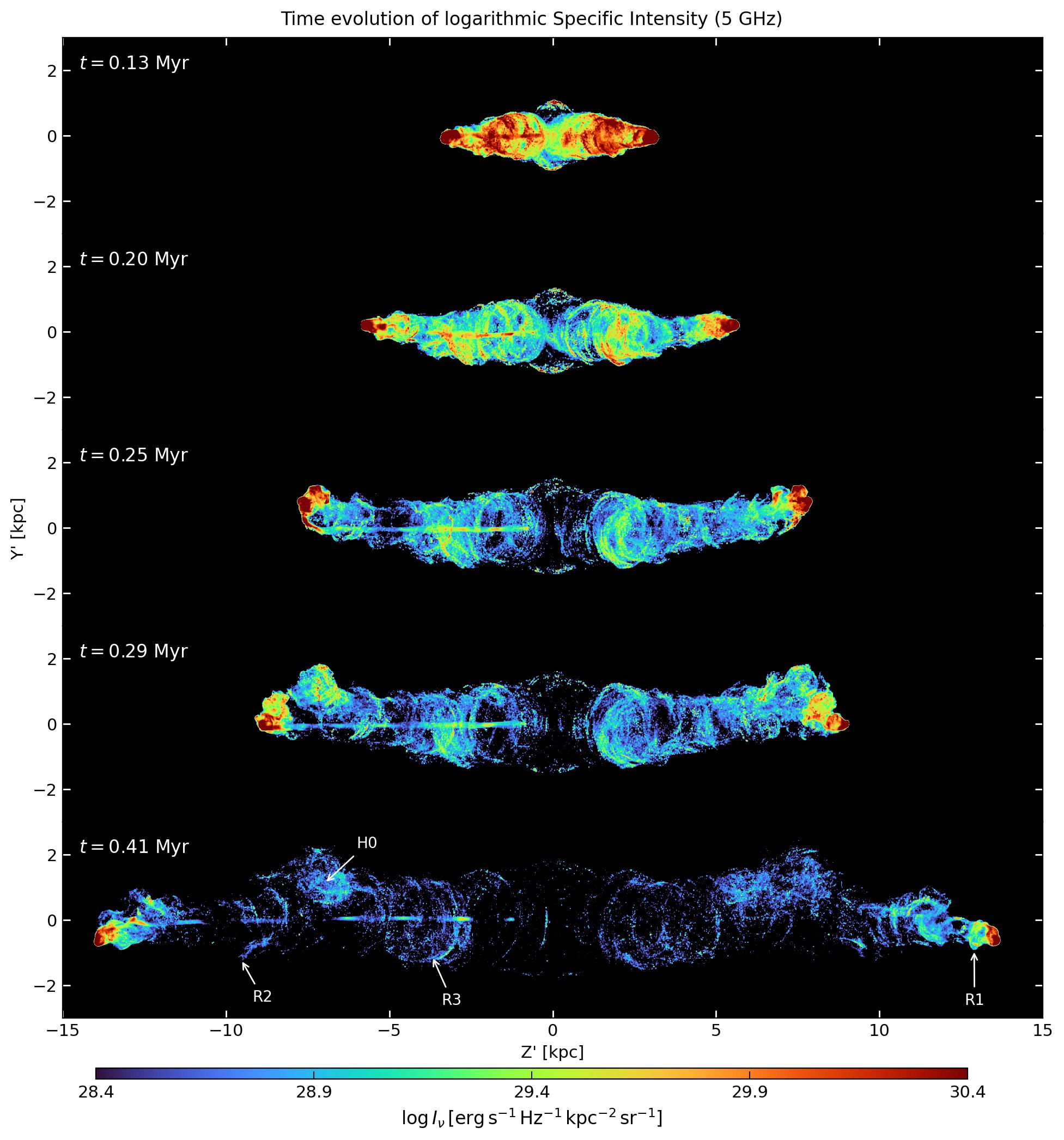}
\caption{Logarithmic surface brightness ($\log I_{\nu} [\flux]$) of J45 on an image plane oriented at $\theta_I = 45^\circ$. The emission is shown at different times for a fixed observed frequency of 5~GHz. The colorbar range is fixed, and regions with values below the lower limit are excluded. R1, R2, and R3 are the arc or ring-like features in the jet's cocoon, and H0 indicates the location of the previous hotspot of the jet.}
 \label{fig:synch_45deg_5ghz}
\end{figure*}

\begin{figure*}
\centering
\includegraphics[width=\linewidth, keepaspectratio]{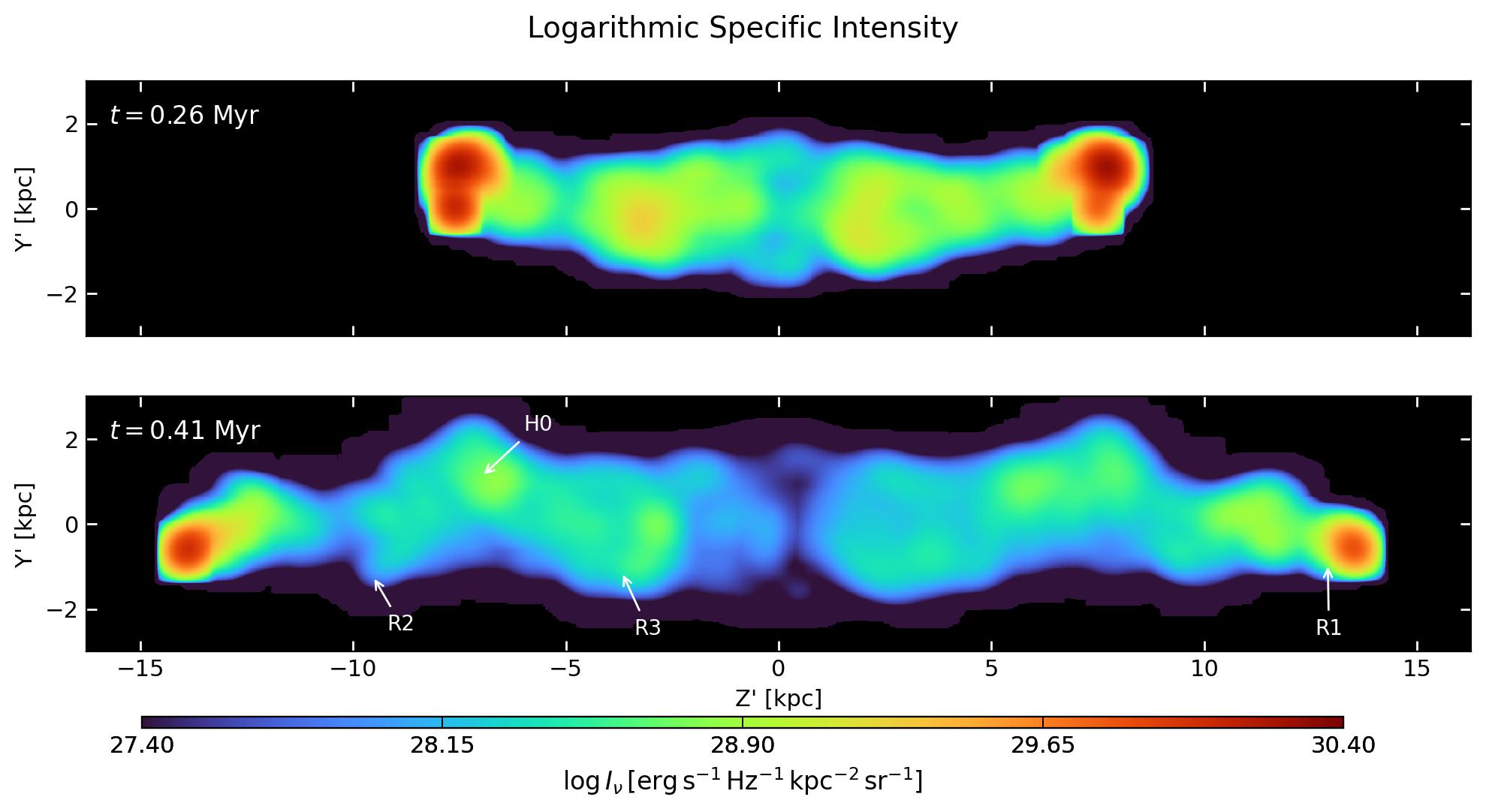}
\caption{Same as Fig.~\ref{fig:synch_45deg_5ghz} at 0.26 and 0.41~Myr, but the specific intensity is smoothed with a Gaussian beam of sigma 300 pc, and truncated at twice this scale. The colorbar range is set to 3-dex to highlight the low-flux regions.}
 \label{fig:synch_45deg_5ghz_smooth}
\end{figure*}

\begin{figure*}
\centering
\includegraphics[width=\linewidth, keepaspectratio]{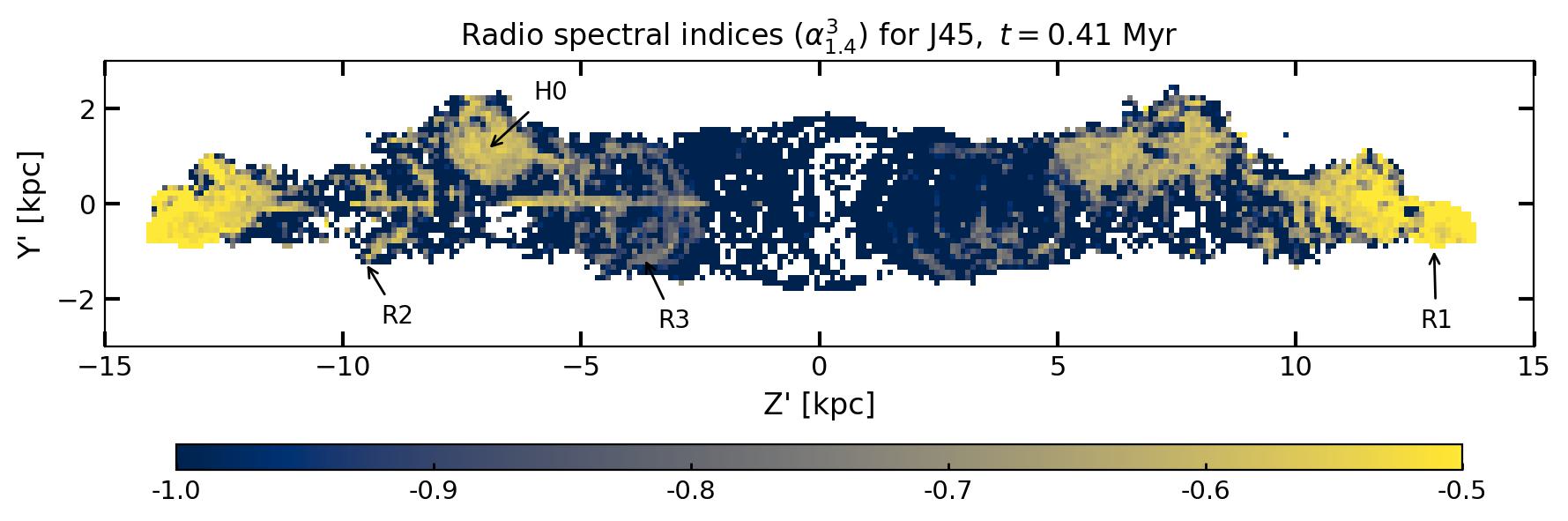}
\caption{Radio spectral indices for a frequency range of 1.4 - 3~GHz in domains of size $100~\pc^2$. The values are estimated for regions with an intensity range used in Fig.~\ref{fig:synch_45_45deg} and~\ref{fig:synch_45deg_5ghz}.}
 \label{fig:spec_45deg_j45}
\end{figure*}

\subsection{Multi-frequency emission from a large-scale jet}
\label{sec:synch_j45}
 
In the previous sections, we examined the synchrotron emission and spectral indices from compact ($Z \leq 4~\mathrm{kpc}$), low-density jets (density contrast $\eta \lesssim 4 \times 10^{-4}$) that exhibited kink instability. In this section, we turn our attention to a larger-scale ($\sim20~\mathrm{kpc}$), higher-power ($10^{45}~\mathrm{erg~s^{-1}}$), referred to as J45 in Table~\ref{tab:sim_table}. The following discussion addresses the corresponding emission morphology and spectral index variations, with particular focus on the maps shown in Figs.~\ref{fig:synch_45_45deg} - \ref{fig:spec_45deg_j45}, which are presented for an inclined line of sight (LOS).

\subsubsection{Morphology of multi-frequency synchrotron emission}

The intensity maps for the large-scale jet (J45) at $\theta_I = 45^\circ$, for a frequency range of 0.15 - 40~GHz, are shown in Fig.~\ref{fig:synch_45_45deg}. The specific intensity limit for the colorbar is kept for 2 dex below the maximum flux attained at 0.15~GHz, and regions below the lower limit are omitted here for comparison. Thus, from this figure, we aim to examine the morphological differences of the large-scale jet at different frequencies, using a fixed flux detection range.

In Fig.~\ref{fig:synch_45deg_5ghz}, the evolution of specific intensity is shown with time for a fixed frequency of 5~GHz. In contrast to the low-power jets, the jet-related features -- namely the bright spine and prominent hotspot -- are clearly visible, as also seen in Paper I. This seems to be facilitated by the greater kink stability of the high-power jet and its narrow cocoon, both of which prevent the cocoon flux from overwhelming the total integrated radio emission, even when the jet is compact. The hotspot remains particularly bright due to the relative absence of frequent kink instabilities, unlike in the low-power cases, where recurrent kinks disrupt and weaken the jet's evolution. It can be seen in Fig.~\ref{fig:visit_vz}; where we show the 3D volume rendering of the $\beta_z = v_z/c$ for J45 (left) and J43 (right) at different times. The comparison demonstrates that J45 propagates steadily with only localized deformations near its head, while J43 develops frequent kink distortions accompanied by pronounced wobbling of the jet head in different directions. Thus, there can be local growth of instabilities of the jet-beam, as we see for J45, while the jet remains straight. Previous studies \citep[e.g.][]{mignone_2010,massagalia_2019,mukherjee_2020} have shown that local kinks can form in the presence of a non-negligible toroidal magnetic field. In the Appendix~\ref{app:kink}, we discuss in detail the numerical details of kink instability and its applicability to J45.

Fig.~\ref{fig:synch_45_45deg} shows that the jet extends $\pm 15~$kpc on the image plane in the $Z-$direction, with resolved hotspots, spine and cocoon features. The left side jet is approaching the observer, and the right one is receding away. At low radio frequencies (0.15 - 1.4~GHz), the emission is widespread across the jet and the surrounding cocoon. The CREs are less affected by synchrotron aging, and lower-energy electrons dominate the emission. At intermediate frequencies (5~GHz), the emission is more concentrated around the jet's spine and near the head. The lobes are fainter and appear more patchy, reflecting moderate synchrotron aging. The flux from the cocoon becomes weak at high radio frequencies, making it below the detectable range. At 20 and 40~GHz, the higher emission is confined to the compact regions along the jet's spine and hotspots. The persistent brightness of terminal hotspots even at 40 GHz suggests ongoing or recent acceleration, indicating that the hotspots are an important site for acceleration in FRII-like jets. It should be noted that since the approaching jet is Doppler boosted, this also enhances the apparent flux from its spine and hotspot. Nevertheless, since the CREs along the jet have higher energy than the surrounding cocoon, they are likely to be seen at higher frequencies where the CREs in the cocoon would have cooled down. Thus, the maps show morphological differences across frequencies, with the high-frequency maps emphasizing localized features like knots or internal shocks, while low-frequency maps show the broader cocoon.

In Fig.~\ref{fig:synch_45deg_5ghz}, we examine the temporal evolution of the observable jet emission over the same detection range as earlier. The lobes surrounding the jet are clearly visible at $5,\mathrm{GHz}$ during the early stages ($t = 0.13 - 0.20\,\mathrm{Myr}$), but they gradually fade with time. A similar trend is seen at the edge-on intensity evolution at $20,\mathrm{GHz}$ in Fig.~\ref{fig:synch_90deg_20ghz}. In both cases, radiative cooling increasingly dominates within the cocoon as the jet enlarges, leading to a progressive decline in its surface brightness. Additionally, rings or arc-like features (R1, R2, and R3) can be seen around the jet spine in Fig.~\ref{fig:synch_45deg_5ghz}. Similar features are also present in the synchrotron flux maps obtained from the post-processing analysis using the local physical variables (not CREs) in Fig.~\ref{fig:synch_63_pp_5ghz}, as described in Paper~I. The enhanced flux in these arcs coincides with regions of increased thermal and magnetic energy associated with the back-flowing plasma near the contact discontinuity (CD) of the cocoon. The regions also seem to align with large-scale Kelvin-Helmholtz vortices that develop along the CD \citep{ratnaparkhi_2026}, which is very common in simulations of jets where the jet injection properties are, on average, symmetric and steady. These features become particularly prominent arc-like for an inclined line of sight. However, in nature such symmetric structures will be rare due to small-scale intermittencies and also jet dynamics. For example, the fluctuating jet simulations \citep{perucho_2023, elley_2026} can disrupt the backflow and potentially weaken or wash out such structures.

However, arc-like structures have been observed in several sources \citep[e.g.][]{kraft_2012,timmerman_2022,rubeis_2025}, but are not very common. Their origin remains uncertain and has been attributed to several possible mechanisms, including multiple jet episodes \citep{timmerman_2022}, non-linear instability modes in jets \citep{perucho_2023}, and the possibility of local annular shocks in the back-flowing plasma \citep{saxton_2002}. However, any localized structures can become less prominent due to resolution effects, making the emission across the lobes appear more homogeneous and diffuse. A similar effect is seen in the smoothed emission maps using a Gaussian beam with a standard deviation of 300 pc, truncated at twice this scale, as shown in Fig.~\ref{fig:synch_45deg_5ghz_smooth}. This corresponds to a resolution of approximately 375 milliarcseconds at a redshift of 0.1, comparable to the resolutions achieved in some previous observational studies \citep{jarvis_2019,jarvis_2021}. The colorbar range is kept fixed at 3 dex to highlight the low-flux regions, as smoothing reduces the overall flux contrast across the lobes. The lobe appears closely homogeneous in radio, and the prominent features, such as the bright spine and arcs, are not clearly distinguishable now.

\subsubsection{Distribution of spectral indices}
In Fig.~\ref{fig:spec_45deg_j45}, the point-to-point radio spectral indices are shown for J45 for $1.4-3~$GHz, and at a viewing angle of $45^\circ$. The general features appear similar to what is seen for the compact jets in Fig.~\ref{fig:spec_45deg_jw}.
The spectral index gradient, from flatter at the jet head to steeper towards the core, confirms radiative aging due to synchrotron losses. The jet spine, hotspots, and nearby regions have flatter spectral indices ($\sim-0.5$), which suggests ongoing or recent particle acceleration. The regions behind the hot spots and toward the core have mixed populations of CREs, with flatter as well as steeper indices ($\alpha \approx -1.1~ {\rm to}~-0.6$). For instance, the flatter regions coincide with H0, which is the site of the previous hotspot (as visible from the third and fourth panels in Fig.~\ref{fig:synch_45deg_5ghz}), and hence the recently accelerated CREs are deposited here. The steep-spectrum near the core is filled with old, synchrotron-aged plasma in the backflow.

The spectral index is flatter in the arc near the jet's head (R1) and gets steeper as one goes near the jet's base (R2, R3). Thus, the variation in spectral index among these features can provide a promising clue to such features, when seen in observations, as discussed in the previous section. However, it should be noted that the sensitivity of the observations is highly dependent on the frequency, which can further limit the observed properties, as we discuss below.

\subsubsection{Observational Limitations \& Relevance:}
\begin{itemize}
  
\item  \textit{Frequency-dependent sensitivity:}

        Low-frequency observations (e.g., using LOFAR (LOw Frequency ARray), MWA (Murchison Widefield Array), GMRT (Giant Metrewave Radio Telescope)) are well suited to detecting diffuse lobe emission, although they typically provide lower angular resolution\footnote{LOFAR-VLBI (Very Long Baseline Interferometer) can attain a resolution of 0.3''.}. High-frequency telescopes (e.g., ALMA (Atacama Large Millimeter Array), VLA at high band) resolve fine features but miss extended, faint lobes. From Fig.~\ref{fig:synch_45_45deg} it is clear that for a fixed detectable intensity range, the extended lobes can be seen at low frequencies, and the emission is limited to the jet's axis and hotspot at higher values. However, additionally, resolution limitations can wipe out the inhomogeneities, and can make the emission more uniform throughout (e.g. see Fig.~\ref{fig:synch_45deg_5ghz_smooth}). Among observations, the ``classical double'' radio jet morphology with bright spine (with ``Doppler'' boosted) ending with bright hotspots and faint lobes around them is seen for a number of sources at low frequencies of around $5~$GHz \citep{bridle_1989, clarke_1992, bridle_1994, schnoemakers_2000}. The cocoon and/or lobes are clearly visible at low frequency but vanish at high frequency, showing how essential low-frequency radio observations are for capturing the spatial extent of the jet's impact on the ambient medium. These sources are extended to several hundred kpc in size, much larger than the J45 jets presented in this paper. 

 \item    \textit{Surface brightness sensitivity at high radio frequencies:}

        Observing faint lobe structures at high frequencies is very challenging due to both synchrotron dimming and telescope limitations. Beam smearing and sensitivity drop can suppress extended emission at high frequency, and thus it is not so common to observe the large-scale jets at very high radio frequencies. Cygnus~A is a clear, well-studied example in the nearby universe where hotspots in~100~kpc jets are observed at high radio frequencies such as $10-40$~GHz \citep{carilli_1999, pyrzas_2015}, and the extended lobes are visible at low radio frequencies around few hundred MHz \citep{lazio_2006, mckean_2016}, analogous to what we see in Fig.~\ref{fig:synch_45_45deg} (also Fig.~\ref{fig:synch_45deg_5ghz_smooth}), and also in several other sources \citep{godfrey_2009, orienti_2020}. Most other FRII galaxies are faint at these frequencies unless they are very nearby or powerful.

\end{itemize}

\begin{figure*}
 \centerline{
\def\arraystretch{1.0}
\setlength{\tabcolsep}{0.0pt}
\begin{tabular}{lcr}
      \includegraphics[width=0.95\linewidth,keepaspectratio]{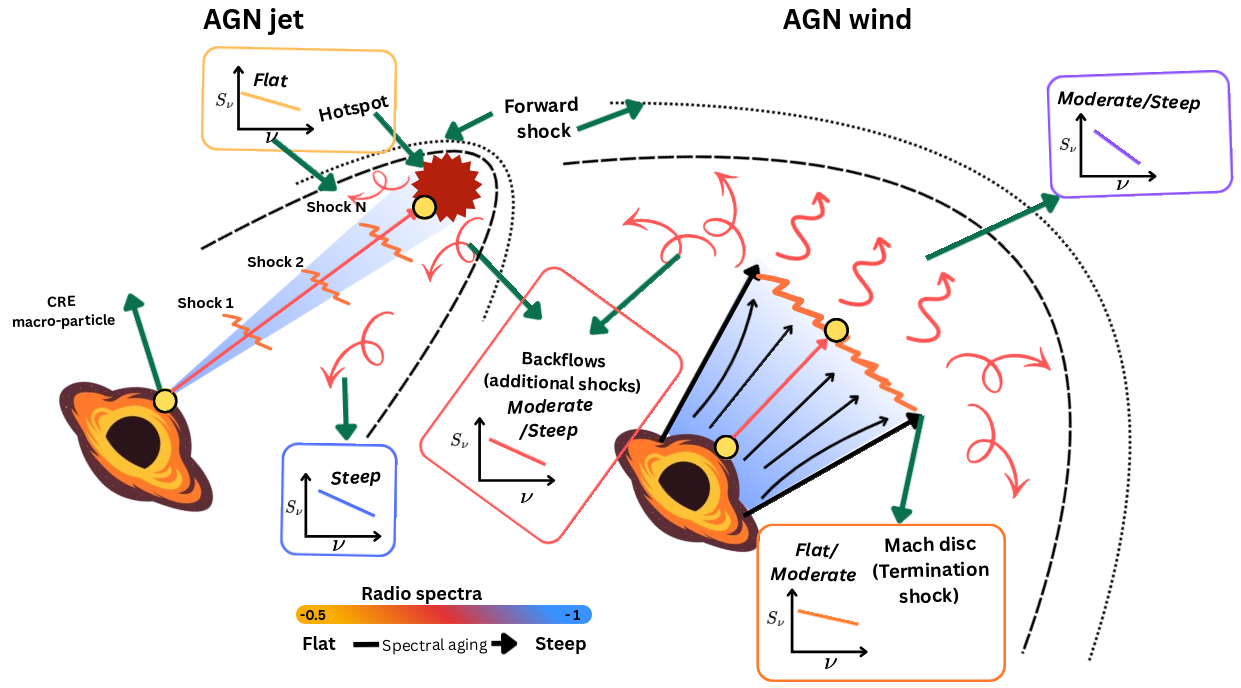}
     \end{tabular}}
\caption{Schematic diagram illustrating the evolution of a CRE macroparticle in a jet (left) and wind (right). The radio spectra originating from different regions are displayed, as indicated by results from this study. The index range is chosen based on the ranges seen in the cocoon from Figs.~\ref{fig:synch_90deg_jw} and~\ref{fig:synch_90deg_jw_hf}. The light winds display steeper indices than the denser winds, and hence the local variations in the wind spectra are indicated.}
 \label{fig:cartoon_jet_wind}
\end{figure*}

\section{Summary and discussion}
\label{sec:discussion}
While relativistic jets are traditionally considered the primary source of AGN radio emission, growing evidence suggests that AGN-driven winds may also contribute significantly \citep{hwang_2018,klindt_2019,morabito_2019,silpa_2020,sargent_2026}. In compact AGN, however, disentangling the relative contributions from jets and winds remains challenging, necessitating robust diagnostics to distinguish between the two mechanisms \citep{njeri_2025,fawcett_2025}. In this context, our previous studies \citep{meenakshi_2023,meenakshi_2024} predicted the general characteristics of synchrotron emission and polarization arising from magnetized jets and winds. These predictions were made in post-process analysis, based on instantaneous physical quantities such as pressure and magnetic fields. In this study, we utilized the LP module \citep{vaidya_2018} in \textsc{pluto} \citep{mignone_2007} to explore multi-frequency emission in kpc-scale jets and winds. The particles are injected into the jets and winds, and their physical properties and spectra are tracked. As a result, the particle distribution function accounts for losses due to adiabatic expansion and radiative losses from synchrotron and inverse-Compton processes (due to CMB radiation). Additionally, the CRE spectra are updated as they cross the shocks, maintaining a record of particle evolution and its impact on current statistics.

We have observed that CREs evolve differently in jets and winds (see Figs.~\ref{fig:lorentz_jw} and~\ref{fig:shkn_jw}), influencing their spectral properties. In jets, particles are primarily accelerated along the spine and at the hotspot \citep[as also shown in][]{mukherjee_2020,dubey_2023}, while in winds, the acceleration happens at the Mach disc and occasionally in the backflows. Our results show that the forward shock produced by jets is stronger than that in winds of comparable power, when compared at similar spatial extents of both. One may argue that the jet simulations have a somewhat higher spatial resolution ($7.8$ pc per cell) than the wind simulations ($10$ pc per cell), which can affect the shock detection in the winds. However, it should be noted that the resolution across the dynamical features is important to detect shocks in the flow, and it is not straightforward to compare jets and the winds as they both have different local shock structures. Nevertheless, as discussed above, most particles in the winds are accelerated at the termination shock near the base, which is well resolved by more than 40 cells across the wind diameter ($400$ pc). Moreover, the sharp decline beyond two shock crossings (Fig.~\ref{fig:pdf_shkn}) indicates that repeated shock encounters contribute only weakly to particle acceleration. Thus, although unresolved internal shocks cannot be entirely excluded, the particle spectral indices are primarily determined by the well-resolved termination shock, and the modest resolution difference is unlikely to significantly affect our main conclusions.

Additionally, since shocks are more frequent in jets than in winds, numerical resolution can play an important role in determining the shock-acceleration histories of CREs and their resulting spectral indices. In Appendix~\ref{app:res_test}, we assess the effects of numerical resolution and find that, although some shock encounters may be missed at lower resolution (i.e., 6 cells across the jet radius employed in this study), the overall spectral-index distributions remain broadly consistent. This suggests that the resolution adopted in our simulations is sufficient to capture the main trends in CRE spectral evolution and that our conclusions are not significantly affected.

Below, we summarize our findings on how particle acceleration in jets and winds shapes the multi-frequency synchrotron emission and radio spectra, and how our results relate to the observations and other studies.

\subsection{Non-thermal radio emission in jets and winds}

    We find that in compact low-power jets (e.g., J43), the radio emission from the cocoon can be dominated by the older CREs, frequently replenished by the backflows from the jet's head (Fig.~\ref{fig:synch_90deg_jw}). Due to several frequent kinks, the jet's spine and hotspot are not clearly detectable in this case. However, for the higher-power jet J44, the hotspot is already prominently detectable at low radio frequencies, and with increasing frequency, the hotspot and nearby regions start to become brighter than the cocoon. Similar trends are also observed at the inclined LOS in Fig.~\ref{fig:spec_45deg_jw}. Thus, in low-power jets, a volumetrically dominant, laterally expanded cocoon can overwhelm the narrow, intrinsically bright spine. 
    
    For high-power, wider, large-scale jet J45, the jetted features (spine and hotspot) are comparatively more discernible than the compact jets, as shown in Fig.~\ref{fig:synch_90deg_20ghz}. However, their prominence evolves over time as the flux from both the jet and the cocoon decreases. This is assisted primarily due to the narrower cocoon compared to the low-power jets and the greater kink stability of J45, which prevents the cocoon emission from overwhelming that of the jet spine, even during the compact phase (see Figs.~\ref{fig:synch_45deg_5ghz} and~\ref{fig:synch_90deg_20ghz}). At inclined view (Figs.~\ref{fig:synch_45_45deg} and~\ref{fig:synch_45deg_5ghz}), the approaching spine is boosted and appears distinctly brighter than the cocoon at low radio frequencies here. In CSS/GPS sources, double-lobe structures with bright hotspots at the ends are common \citep{orienti_2016}, as we have seen for the compact jets. However, jet-like collimated features or bubbles, and sometimes hotspots, are also observed in small-scale radio sources in systems, such as NGC 6764 \citep{kharb_2010}, NGC 1068 \citep{mutie_2024}, and NGC 4151 \citep{williams_2017}, to name a few.

   On the other hand, in winds, the bright arcs near the cocoon head and the regions above the Mach disc dominate the emission at lower radio frequencies (Figs.~\ref{fig:synch_90deg_jw}). These bright arcs, which result from the top edges of the lateral streams emerging from the Mach disc, manifest as bright circular arcs at inclined viewing angles. In dense winds, the Mach disc appears to be the brightest and wider when compared to light winds. This emission appears as wide, bright circular arcs at inclined LOS. Observationally, radio detections of suspected AGN-driven winds are rare, and when observed, the emission is typically compact \citep{peng_2020, yang_2022, calistro_2024}. It is often non-trivial to characterize the radio-emission properties of compact sources (see \citet{panessa_2019}, \citet{harrison_2024}, for observational and \citet{mukherjee_2025} for simulation-focused reviews). However, combining multiple diagnostics -- such as emission, polarization, and spectral features -- can provide valuable guidance for identifying their observational signatures, as demonstrated in our studies.

    \subsection{Radio spectral indices in jets and winds}
    
    In all the jet cases, a common trend is that the radio spectra get steeper from the hotspot towards the lobes (see Fig.~\ref{fig:synch_90deg_jw}). Such a feature has been observed in the kpc-scale jets \citep{mutie_2024,orienti_2004,orienti_2012}, and also in large-scale FRII jets \citep{kharb_2008,vaddi_2019,baghel_2023}. These studies find spectral indices of around -0.5 at the hotspot, and steeper in the lobes ($\alpha\sim-1$) at GHz frequencies, similar to the values we obtained here. Additionally, \citet{murgia_2003} found that spectral break frequency decreases from the hotspot towards the core, indicating enhanced spectral aging as one moves from the hotspots towards the galaxy's core \citep{carilli_1991}. In our simulations, the flatter spectral indices ($\alpha \gtrsim -0.6$) observed in the upper parts of the cocoon, and steeper downwards, similarly suggest the presence of newly accelerated CREs near the jet's head.

  For winds, the radio spectra become steeper as one moves away from the Mach disc (see Fig.~\ref{fig:synch_90deg_jw}). At lower radio frequencies ($1.4-3~$GHz), the spectral indices from regions above the Mach disc have comparable values in light wind cases (W43/W44-light). The dense winds show comparatively shallow values ($\alpha\approx -0.65$) at the Mach disc. \textit{These values are comparatively steep than the spectral indices obtained from the upper regions of the jets of similar power.} In agreement with \citet{dubey_2024}, this suggests that particle acceleration is governed more by the number of shocks encountered than by their individual strength. However, at higher frequencies ($3-10~$GHz), flatter spectral indices are observed only near the Mach disc, and there is a sharp lowering in the lateral regions and close to the cocoon head. This is because \textit{CREs undergo radiative losses ahead of the Mach disc in winds, and hence, lead to steeper ($\alpha \lesssim -1.2$) spectra in these regions (see Fig.~\ref{fig:synch_90deg_jw_hf}). In contrast, in compact jets of similar scales, the spectral slopes across the cocoon remain nearly similar up to high radio frequencies} (Fig.~\ref{fig:synch_90deg_jw_hf}). 
  
We found that in winds, the forward shock is strong when they are compact ($\lesssim 1$~kpc) but loses its strength as the wind propagates to larger distances (4~kpcs). This results in a steeper spectral index from SAM in winds compared to jets in our study. \textit{The Mach disc in winds is found to be more efficient than the forward shock at accelerating particles to high energies.}

The mean values listed in Table~\ref{tab:index_table_2} indicate that winds  -~particularly light winds - tend to exhibit steeper spectral indices than jets. Here, the indices typically range from -1.0 to -0.9, whereas in jets, they are around -0.6. However, one should note that the density of the wind can change these integrated measurements, as we find that a dense wind (e.g., W43-dense) can show values closer to a jet. Hence, in such cases, spatially resolved indices are a more powerful diagnostic. Moreover, in low-power kink-unstable jets (J43 here), the mean spectral index (i.e., -0.85) is quite steeper than J44 (-0.59), as newly injected CREs may not dominate the cocoon’s emission. However, when limited to regions near the jet's head, we observe a comparatively flat mean index of around -0.78. Similar trends are seen in the integrated radio SEDs in Fig.~\ref{fig:tot_spectra}.

Theoretical studies have also investigated radio emission arising from particle acceleration in winds, focusing on both thermal and non-thermal components \citep[e.g.][]{nims_2015}. In particular, \citet{nims_2015} showed in their model that steep radio spectra are expected at frequencies where CREs have radiated away most of the energy gained at shocks\footnote{$\alpha\approx-1$ for CREs with a power-law index of $\delta=2$, and steeper for larger $\delta$.}. Consistent with this, we also obtain steep spectral indices ($\alpha \lesssim -1$), especially in the light-wind cases. This steepening arises from several factors: the termination shocks are weaker in light winds than in dense winds, and the slow propagation, together with weak shocks within the wind cocoon, allows sufficient time for significant radiative cooling of CREs. Collectively, these effects can lead to systematically steeper mean spectral indices in the wind cases. To summarize the above discussion, we present a schematic illustration in Fig.~\ref{fig:cartoon_jet_wind}, showing the evolution of a CRE (represented by the yellow dot) as it propagates through AGN jets and winds and the resulting variation in its radio spectrum. The red curves trace the trajectory of the CRE, with the backflows indicated by the curved streamlines. In addition to the major shocks encountered by the CRE along the jet (orange ragged lines) and the Mach disk shock in the wind, the backflowing plasma also hosts some additional shocks. However, many of the CREs undergo cooling in the cocoon in our simulations, which gives rise to low Lorentz factors in these regions (see Fig.~1); however, the shocks here are expected to be shaped by the interaction of AGN-driven outflows with the inhomogeneous multiphase medium \citep[as explored in our previous studies,][]{mandal_2021,meenakshi_2022a,meenakshi_2022b}, which can affect the emission characteristics arising from shocked CREs \citep[see][for a recent work]{almeida_2026}. The corresponding radio spectra produced at these different locations are also indicated in the schematic. This diagram was created using the online graphic design platform Canva.\footnote{https://www.canva.com}

\subsection{Limitation and future aspects}
\label{sec:limitation}
In this section, we discuss some limitations of our study that need further investigation for a more accurate comparison with observations. Compact jets and winds are expected to interact with the host galaxy’s clumpy, multi-phase ISM at kpc scales \citep{Mukherjee_2016,Cielo_2018,meenakshi_2022a,meenakshi_2022b}, which can produce additional thermal and non-thermal emission. Additionally, several factors can also influence synchrotron emission at different frequencies, including synchrotron self-absorption (SSA), free-free absorption (FFA), and inverse-Compton cooling due to AGN radiation. At low radio frequencies, synchrotron self-absorption (SSA) can flatten the observed spectrum by rendering the plasma optically thick to its own emission. This effect reduces the observed flux at low frequencies and can produce a characteristic spectral index of 2.5 \citep{rybicki_1979}.

Synchrotron self-absorption (SSA) is expected to be particularly important in compact radio sources, where high electron densities and strong magnetic fields can suppress low-frequency emission \citep{odea_1998, odea_2021}. Since our simulations begin at a scale of 100 pc, the compact AGN core, where SSA is typically strongest, is not resolved. Therefore, the very flat spectral indices seen in the `true cores' of jetted systems are not resolved in our study. Within the kpc-scale lobes, SSA, if present, would most likely arise from the densest and most strongly magnetized plasma-dominated regions, particularly near the jet head or Mach disc in the winds. Thus, these regions can enhance SSA locally. We find maximum magnetic field strengths of 0.3, 0.06, and 0.11 mG for J43, W43-light, and W43-dense, respectively, for the jet/wind plasma-dominated regions. A similar trend is seen in the higher-power cases J44 and W44-light, which reach maximum fields of 0.854 and 0.13 mG, respectively.

The SSA optical depth scales with the absorption coefficient; therefore, for comparable CRE number densities, the strength of SSA depends primarily on the underlying magnetic field strength and the electron energy distribution index as \citep{rybicki_1979}:

\begin{equation*}
    \alpha_{\nu} \propto B^{(\delta + 2)/2} \, \nu^{-(\delta+4)/2}
\end{equation*}

The spectral indices in these compact shock regions do not differ substantially, with jets J43 (J44) reaching maximum values of $\alpha \approx -0.6$ ($-0.5$), while light winds typically show $\alpha \approx -0.75$ at the Mach disc (Fig.~\ref{fig:synch_90deg_jw}). Consequently, the absorption coefficient is expected to scale mainly with the magnetic field strength (approximately $\propto B^{2}$ for $\delta \sim 2$–3), leading to orders-of-magnitude increase in the coefficient in jets compared to winds. Thus, the jets are expected to be more susceptible to the localized SSA effects. A detailed analysis of the impact of SSA on the SEDs is deferred to future work.

Simulations by \citet{bicknell_2018} indicate that FFA by the ISM can produce even steeper radio spectra than predicted by SSA alone, a phenomenon observed in several sources \citep{hurley_2017, callingham_2017}. Additionally, inverse-Compton cooling \citep{nims_2015,yamada_2024} can have a significant impact on the radio spectra for highly luminous AGN, adding further complexity to the scenario. Our current study focuses on a simplified case, examining emissions solely from jets and winds, assuming optical thin synchrotron emission, and without accounting for ISM contamination. A systematic exploration including different physics discussed above and their effects on the observed synchrotron emission from compact jets and winds is reserved for later study.

\section{Conclusions}

\label{sec:conclusions}
In this work, we investigated the multi-frequency synchrotron emission and radio spectral properties of kpc-scale AGN jets and winds using self-consistent particle evolution with the LP module \citep{vaidya_2018} in \textsc{pluto} \citep{mignone_2007}. Unlike earlier post-processed approaches (Papers I and II), the present study tracks CRE injection, shock re-acceleration, and energy losses due to adiabatic expansion, synchrotron, and inverse-Compton processes. This enables a more realistic comparison of how particle acceleration and transport shape the observable radio morphologies and spectra in compact jets and winds of comparable power and spatial extent. Below, we list the main conclusions from our analysis.

\begin{itemize}

\item \textbf{Distinct particle acceleration sites in jets and winds:}
CREs in jets are primarily shock accelerated along the spine and at the hotspots, whereas in winds acceleration occurs at the Mach disc, and occasionally in backflows (as inferred from Figs.~\ref{fig:lorentz_jw} and~\ref{fig:shkn_jw}). For comparable power and spatial extent, jets produce narrower and stronger forward shocks than winds. In winds, the forward shock weakens rapidly as the outflow expands, while the Mach disc remains the dominant and persistent particle acceleration site.

\item \textbf{Radio emission morphologies in jets:}
Low-power compact jets (J43 and J44) can exhibit cocoon-dominated radio emission extending to relatively high radio frequencies (left panels in Figs.~\ref{fig:synch_90deg_jw} and~\ref{fig:spec_45deg_jw}). This behaviour is driven by the continuous replenishment of CREs through backflows from the jet head, which sustains synchrotron emission within the cocoon. In contrast, the high-power jet (J45) displays more prominent hotspots and jet spines, particularly when viewed at inclined lines of sight (Figs.~\ref{fig:synch_90deg_20ghz} and~\ref{fig:synch_45deg_5ghz}).

\item \textbf{Radio emission morphologies in winds:}
In winds, the brightest radio emission typically arises from CREs accelerated at the Mach disc (middle and right panels in Figs.~\ref{fig:synch_90deg_jw} and~\ref{fig:spec_45deg_jw}). Light winds (W43/W44-light) show significant contributions from the upper edges of the lateral streams emerging from the Mach disc, which advect CREs toward the cocoon head. These regions appear as bright overhead columns with horizontal arc-like features at $\theta_I=90^\circ$, and as bright circular arcs at inclined viewing angles. In contrast, dense winds (W43-dense) exhibit radio emission dominated by the Mach disc, appearing as wide bright arcs at inclined LOS.

\item \textbf{Spectral index gradients in jets:}
Jets consistently show flat spectra near the hotspot ($\alpha \sim -0.5$ to $-0.6$) that steepen toward the lobes and cocoon (e.g. Figs.~\ref{fig:synch_90deg_jw} and~\ref{fig:spec_45deg_jw}), reflecting spectral aging and the transport of CREs away from acceleration sites. It suggests that even when classical jet features such as the hotspot or spine are not prominently visible in compact jets, the spatial distribution of spectral indices remains a robust diagnostic of jet activity.

\item \textbf{Spectral index gradients in winds:}
In winds, spectral indices progressively steepen with increasing distance from the Mach disc; which is observed in both light and dense wind cases (see Fig.~\ref{fig:synch_90deg_jw}). As one goes to high radio frequencies, only regions very close to the Mach disc retain comparatively flat spectra, while lateral and upper cocoon regions start to exhibit significant steepening due to radiative losses (see Fig.~\ref{fig:synch_90deg_jw_hf}).

\item \textbf{Spectral differences between jets and winds:}
The flux-weighted mean spectral indices in light winds (W43/W44-light: $-1.0$ to $-0.9$) are significantly steeper than the stable compact jets (J44: $\sim -0.6$), as can be inferred from Table~\ref{tab:index_table_2} and radio SEDs in Fig.~\ref{fig:tot_spectra}. This is attributed to the slower propagation speeds, weaker shocks, and longer radiative cooling timescales in these winds. However, the kink-unstable low-power jet (J43) shows a steeper mean index ($\sim -0.85$), largely due to dominant emission from the central and lower cocoon, yielding values comparable to those of winds. Dense stable winds exhibit a comparatively flatter mean index ($\sim -0.7$) than the light winds of similar power.

\end{itemize}

Overall, these results will be key for helping guide and interpret observations. Sufficient resolution in observations may help distinguish morphological characteristics from jets and winds, but as winds and low-power jets are often compact, this may require targeted observations using very long baseline interferometry (VLBI) techniques. Large surveys like FIRST \citep{becker_1995}) or LoTSS \citep{shimwell_2026} have typical resolutions of a few arcsec. The different spectral behaviour between jets and winds may be extremely useful for helping distinguish between compact jets and winds.

\section*{Acknowledgements}
We thank the anonymous referee for useful and constructive comments that helped in improving the clarity of the manuscript. MM acknowledges support by the European Research Council under ERC-AdG grant PICOGAL-101019746, and support by the DFG Research Unit FOR-5195. DM acknowledges the Indo-Italian mobility grant INT/Italy/P-37/2022 (ER) (G). CMH acknowledges funding support from the United Kingdom Research and Innovation (project UKRI2730). LKM is grateful for support from a UKRI FLF [MR/Y020405/1] and LOFAR-UK via STFC [ST/V002406/1]. PK acknowledges the support of the Department of Atomic Energy, Government of India, under the project 12-R\&D-TFR-5.02-0700.  SS acknowledges funding from ANID through Fondecyt Postdoctorado (project code 3250762), Millennium Nucleus NCN23\_002 (TITANs), and Comit\'e Mixto ESO-Chile. The authors (MM, DM, GB, PR) acknowledge support by the Accordo Quadro INAF-CINECA 2017 for the availability of high-performance computing resources. MM gratefully acknowledges the computing time granted by the Resource Allocation Board and provided on the supercomputer Emmy/Grete at NHR-Nord@Göttingen as part of the NHR infrastructure. The resolution test calculations were conducted with computing resources under the project bbp00091.

\section*{Data Availability}
The data underlying this study will be shared on reasonable request to the corresponding author.

\bibliographystyle{mnras}
\bibliography{references}

\appendix

\section{Kink stability analysis for J45}
\label{app:kink}
We find that the J45 jet is mostly straight, however developes local kink-like features near its head at later stages (as can be seen from Fig.~\ref{fig:synch_45deg_5ghz}). Overall, the jet's spine remains straight and is not disrupted by the kink modes. It is known that local relativistic flows in jets can suppress kink modes. For example, Eq. 39 of \citet{bodo_2013} shows a scaling inversely proportional to the fourth power of the jet's Lorentz factor. Although the jets in our study have relatively weak magnetisation, such results hint towards the qualitative trends that at the initial stages in the highly relativistic beam, the local modes are suppressed. However, they manifest later towards the jet head when the jet beam slows down at the site of interaction with the ambient medium.
This behaviour is consistent with the external kink instability discussed by \cite{bromberg_2016}, which primarily affects the jet head. From Eq. (2) in their paper, the characteristic growth time of the instability in the lab frame is approximately the Alfv\'en crossing time around the jet circumference,

\begin{equation}
t_{\rm kink} \approx
\frac{2\pi R_j \gamma_j}{v_A}
\frac{b_p}{b_\phi},
\end{equation}
where $R_j$ is the jet radius, $\gamma_j$ is the bulk Lorentz factor, $b_p$ and $b_\phi$ are the poloidal and toroidal magnetic field components in the fluid frame, and $\varv_A$ is the Alfv\'en speed.
The global kink modes typically require up to 10 growth times to produce substantial jet deformation. The time available for the instability to develop is the dynamical time,
\begin{equation}
t_{\rm dyn} \approx
\frac{L_j}{c\beta_j},
\end{equation}
where $L_j$ is the instantaneous jet height at the given snapshot and $c\beta_j$ is the bulk flow speed. For moderately magnetized jets, the corresponding stability parameter is (using Eq. 29 in their work for less magnetized jets and not assuming equipartition in the magnetic components)
\begin{equation}
\Lambda =
\frac{10t_{\rm kink}}{t_{\rm dyn}}
=
\frac{20\pi R_j \gamma_j \beta_j c}
{v_A L_j}
\frac{b_p}{b_\phi}.
\end{equation}

Jets are expected to become globally kink unstable when $\Lambda \lesssim 1$. We evaluate $\Lambda$ as a function of jet height for the J45 simulation at different times, which is estimated using the mean value of $t_{\rm kink}$ evaluated at the jet's sheath for a given epoch. The magnetic field components and Alfv\'en speed are measured in regions defined by tracer values $0.85 \le {\rm tr} \le 0.95$, where the toroidal field is calculated as $b_\phi=(\bm{b}_x^2+\bm{b}_y^2)^{1/2}$ and the poloidal field as $b_p=|\bm{b}_z|$ in the fluid frame, where $\bm{b}_{x,y,z}$ are the magnetic field components along the $x,y$ and $z$ axis. Although the local ratio $b_p/b_\phi$ spans a wide range ($0.01$--$200$), its mean value remains between $0.3$ and $0.9$.

\begin{figure}
    \centering
    \includegraphics[width=0.75\linewidth]{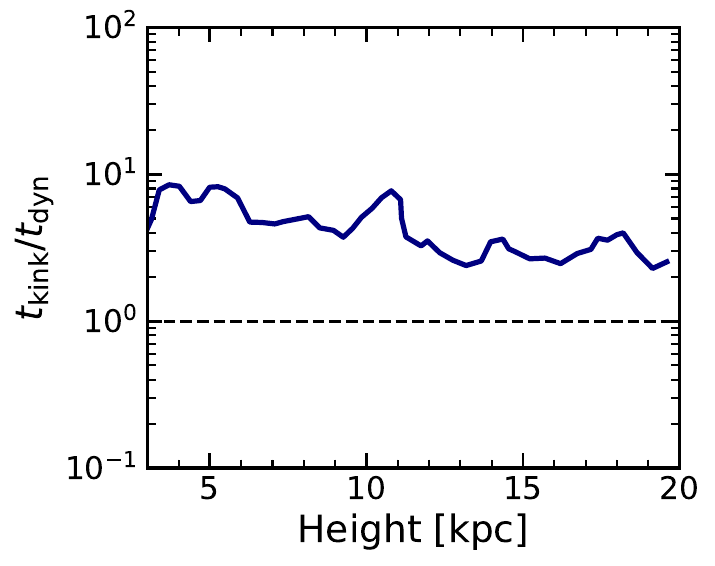}
    \caption{Kink stability parameter ($\Lambda$) as a function of jet's height for the J45 case.}
    \label{fig:kink_param}
\end{figure}

The resulting evolution is shown in Fig.~\ref{fig:kink_param}. We find that $\Lambda$ is close to 10 at earlier times, and saturates at nearly 3 by the end time, i.e., well above the instability threshold. Similar calculations have been performed in previous studies, such as \citet{mizuno_2009, hu_2026}, however, the jets in their studies are more highly magnetized than in our study. Therefore, while localized external kinks can develop near the jet head, they do not have sufficient time to grow into global modes capable of disrupting the jet. This explains why the J45 jet maintains a nearly straight spine, in contrast to the strongly kinked morphology of the low-power J43 (as shown in Fig.~\ref{fig:visit_vz})

For J43, a simple estimate can help explain the strong kink distortions and pronounced bending of the jet spine. According to Eq.~39 of \citet{bodo_2013}, the growth rate of the current-driven kink instability scales as $\omega \propto P_c^{-3}\gamma_c^{-4}$, where $P_c$ is the magnetic pitch and $\gamma_c$ is the Lorentz factor at the jet axis. Assuming comparable magnetic pitch and other jet properties, the lower injection Lorentz factor of J43 ($\gamma_c=2$), compared to J45 ($\gamma_c=5$), implies a kink growth rate approximately $(5/2)^4 \simeq 39$ times higher. The kink instability parameter, which is inversely proportional to the growth rate, is estimated to be approximately 10 for J45 at a jet height of 3 kpc (Fig.~\ref{fig:kink_param}), which is comparable to the height reached by J43 at the end of the simulation (Fig.~\ref{fig:visit_vz}). Applying the same scaling gives an estimated value of $10/39 \simeq 0.26$ for J43, well below the stability threshold. Although this estimate neglects possible differences in the jet structure and magnetic field configuration, it suggests that the lower Lorentz factor of J43 can substantially enhance kink growth, consistent with the strong distortions observed in the simulation.

\section{Additional Plots}
\label{app:j45_plots}
\begin{figure*}
\centering
\includegraphics[width=0.8\linewidth, keepaspectratio]{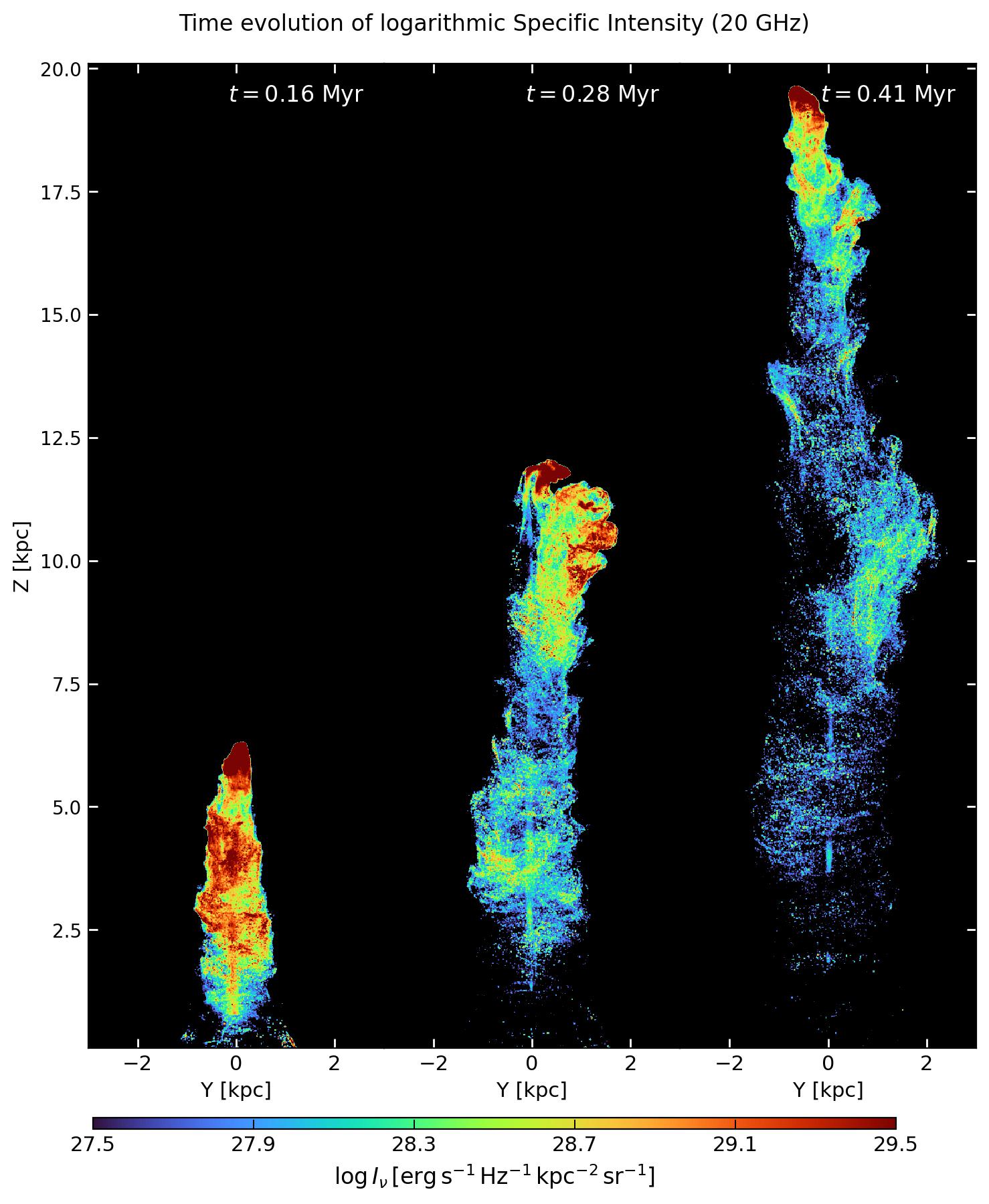}
\caption{Logarithmic specific intensity ($\log I_{\nu} [\flux]$) for J45 on the $Y-Z$ image plane. The emission is shown at different times for a fixed observed frequency of 20~GHz. The colorbar range is fixed to 2-dex, and regions with values below the lower limit are excluded.}
 \label{fig:synch_90deg_20ghz}
\end{figure*}

Fig.~\ref{fig:synch_90deg_20ghz} shows surface brightness maps for the high-power jet model J45 on the $Y$–$Z$ image plane ($\theta_I=90^\circ$) at various times. The color scale spans a fixed range of 2-dex across all times to enable consistent comparison of brightness evolution.

\begin{figure*}
\centering
\includegraphics[width=\linewidth, keepaspectratio]{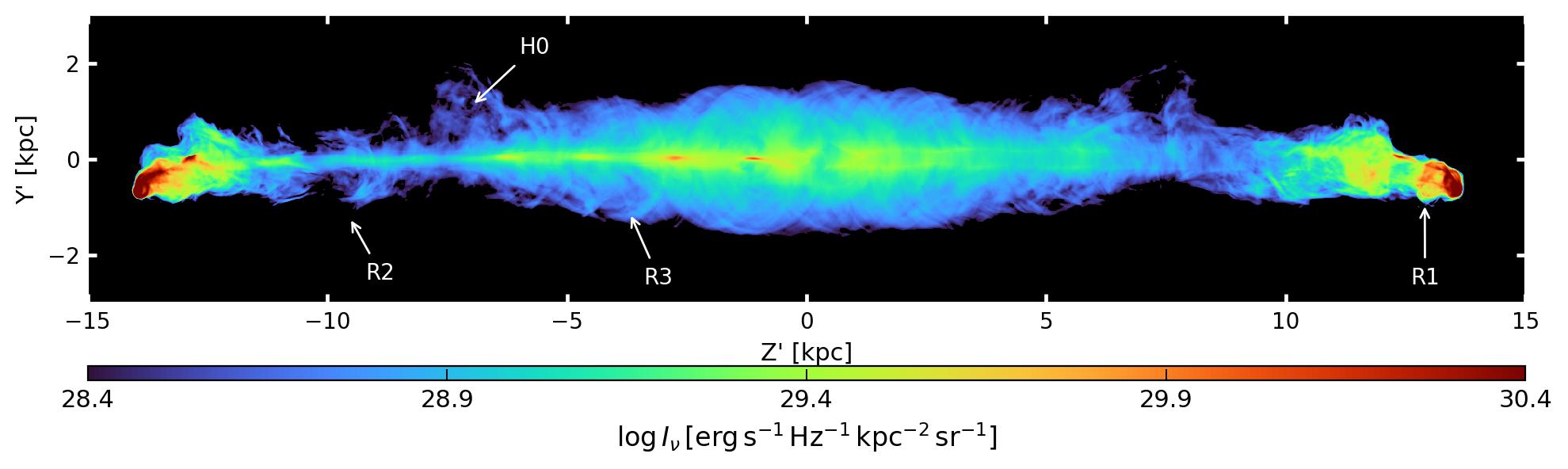}
\caption{Synchrotron intensity at 5~GHz for J45 ($t=0.41~{\rm Myr}$), evaluated in post-process using physical variables, as done in \citet{meenakshi_2023}. This figure corresponds to the lowermost panel presented in Fig.~\ref{fig:synch_45deg_5ghz}, where the left-side jet is approaching the observer.}
 \label{fig:synch_63_pp_5ghz}
\end{figure*}

In Fig.~\ref{fig:synch_63_pp_5ghz}, the synchrotron intensity at 5~GHz, computed in post-processing, is shown for J45 at an inclination angle of $\theta_I = 45^\circ$ at $t = 0.41~{\rm Myr}$. Arc-like features, labeled R1, R2, and R3, as well as the location of the previous hotspot (H0), are marked, consistent with the labeling used in Fig.~\ref{fig:synch_45deg_5ghz}.

\begin{figure*}
\centering
\includegraphics[
    width=0.99\linewidth,
    keepaspectratio
]{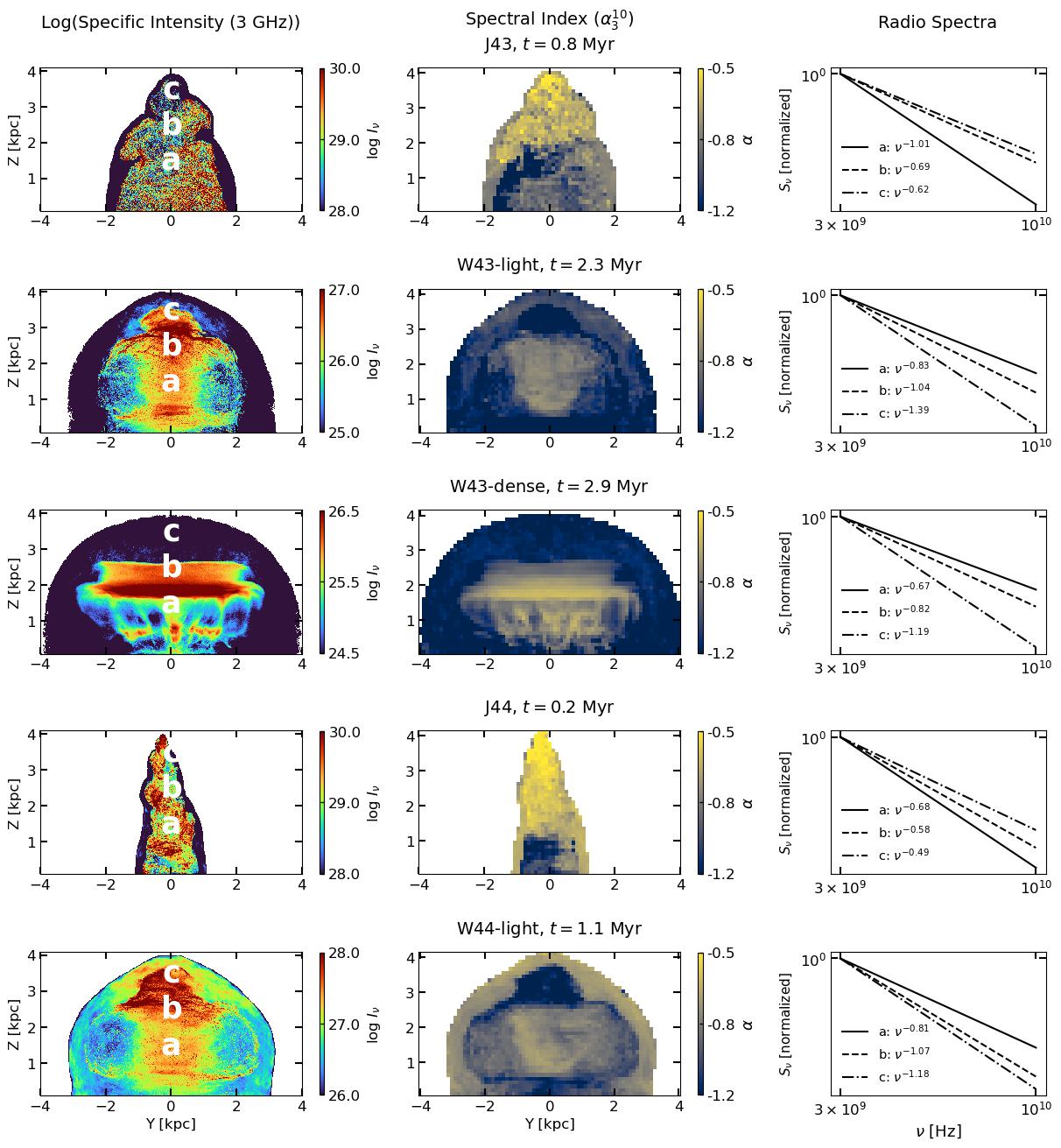}
\caption{Same as Fig.~\ref{fig:synch_90deg_jw}, but the left column show plots for 3~GHz, and the spectral indices and radio spectra are shown for $3-10~$GHz.}
\label{fig:synch_90deg_jw_hf}
\end{figure*}

Fig.~\ref{fig:synch_90deg_jw_hf} presents a set of diagnostic plots for the compact jet and wind models analogous to those shown in Fig.~\ref{fig:synch_90deg_jw} in the main manuscript. From left to right, the panels display: the logarithmic synchrotron flux ($\log I_{\nu}$ [$\flux$]), at 3~GHz; spectral index maps over the 3-10~GHz range; and spatially resolved radio spectra for regions a, b, and c (as marked in the leftmost panels), for both jet and wind cases. All panels are shown in the $Y$–$Z$ image plane ($\theta_I = 90^\circ$).

\section{Resolution study}
\label{app:res_test}

We perform a resolution study for a jet with a power of $10^{44}~{\rm erg\,s^{-1}}$, analogous to the J44 simulation presented in this paper, but within a smaller computational domain of size $(2~{\rm kpc})^3$. Similar to the simulations performed in our study, we employ the piecewise parabolic method (PPM) reconstruction \citep{colella_1984}, the HLLD Riemann solver \citep{mignone_2009}, and RK3 time-stepping. In regions around the injection domain, i.e., within a 200 pc cubical region, we switch to HLL if the pressure ($p$) jump exceeds 4 (i.e., $\delta p/p_{\rm min}>4$), for numerical stability. We consider three grid resolutions of $256^3$, $400^3$, and $600^3$ cells, corresponding to the spatial resolutions of $7.8$, $5$, and $3.33$ pc\footnote{For this case, we switch to HLL when the pressure jump condition ($\delta p/p_{\rm min}>4$) is satisfied in the whole domain, after the jet reaches around 1 kpc, for numerical stability at later stages.}, respectively. Accordingly, the number of cells ($N$) across the jet radius increases from $6$ to $10$ and $15$, with $N=6$ corresponding to the resolution adopted in our main study. Two CRE macroparticles are injected into each computational cell inside the jet injection radius at each time step. The analysis was performed when the jet reached a height of approximately $2$ kpc, at $0.11$ Myr.

\begin{figure*}
    \centering
    \includegraphics[width=0.8\linewidth, keepaspectratio=True]{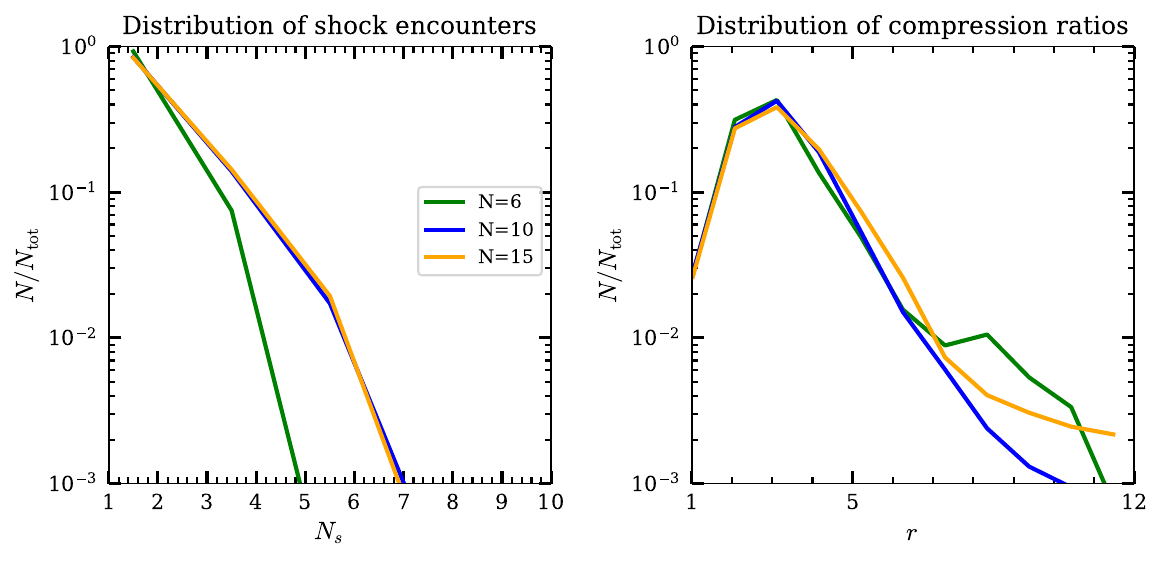} 
    \caption{Distributions of shock encounters and compression ratios for the resolution study of the J44-like run, at a time of 0.11~Myr. $N_{\rm tot}$ indicates the total number of shocked CREs in the simulations. The tests are performed by varying the number of computational cells across the jet radius, i.e., 6, 10, and 15.}
    \label{fig:pdf_shk_cr}
\end{figure*}

Fig.~\ref{fig:pdf_shk_cr} shows the distributions of shock encounters experienced by the CREs (left panel) and the corresponding compression ratios (right panel). The distributions of shock encounters seem to converge after $N=10$, although it might be possible that some shocks are missed for $N=15$ at later stages. In contrast, the distributions of compression ratios remain broadly consistent across the three resolutions for most of the CREs. The distributions show minor differences (which are important for less than 1\% of the shocked CREs), suggesting convergence for the compression ratios. We note that the compression ratio recorded for each CRE corresponds to its most recent shock encounter, rather than its complete shock-crossing history. The differences in the compression-ratio distributions are more apparent at $r>6$, which exceeds the maximum compression ratio for a strong shock (which is 4) for an ideal gas, and may arise from numerical effects. For such cases, we adopt a DSA spectral slope of $-2.23$, corresponding to the ultra-relativistic limit predicted by \citet{kirk_2000}. In contrast to \citet{dubey_2023} (their Fig.~B2), we find that around 10 cells across the jet's radius are sufficient for converging properties from the shocked CREs. This is attributed to our use of PPM reconstruction, the HLLD Riemann solver, and RK3 time-stepping, when compared to the linear reconstruction and HLL solver employed by \citet{dubey_2023}. In particular, PPM provides higher-order spatial reconstruction, while HLLD is less diffusive than HLL in resolving MHD discontinuities. Although some repeated shock encounters can remain resolution-dependent, we have verified that the overall distribution of CRE spectral indices within the jets is similar across different resolutions, suggesting that our main conclusions are reasonably robust against changes in numerical resolution.

\label{lastpage}
\end{document}